\documentclass{article}[12pt,a4paper]
\pdfoutput=1
\usepackage{graphicx}  
\usepackage{dcolumn}   
\usepackage{bm}        
\usepackage{amssymb}   
\usepackage[]{amsmath,amssymb}
\usepackage{graphics,epsfig}
\usepackage[latin1]{inputenc}
\usepackage[T1]{fontenc}
\usepackage{amsfonts}
\usepackage{latexsym}
\usepackage[english]{babel}
\newcommand{\be}{\begin{equation}}
\newcommand{\ee}{\end{equation}}
\newcommand{\beq}{\begin{equation}}
\newcommand{\eeq}{\end{equation}}
\newcommand{\bea}{\begin{eqnarray}}
\newcommand{\eea}{\end{eqnarray}}
\newcommand{\nn}{\nonumber}
\def\be{\begin{equation}}
\def\ee{\end{equation}}
\def\ba{\begin{eqnarray}}
\def\ea{\end{eqnarray}}

\begin{document}

\title{Thermoelectric Transport Coefficients from Charged Solv and Nil  Black Holes }

\author{Ra\'{u}l E. Arias$^{a}$\footnote{rarias@fisica.unlp.edu.ar}, Ignacio Salazar Landea$^{b}$\footnote{peznacho@gmail.com}}
\maketitle

\begin{center}

{\sl $^a$ Instituto de F\'{\i}sica de La Plata - CONICET\\
C.C. 67, 1900 La Plata, Argentina.}

~

{\sl $^b$ Centro At\'omico Bariloche,\\
8400-S.C. de Bariloche, R\'{\i}o Negro, Argentina}

\end{center}

~

~

\begin{abstract}

In the present work we study charged black hole solutions of the Einstein-Maxwell action that have Thurston geometries on its near horizon region. In particular we find solutions with charged Solv and Nil geometry horizons. We also find Nil black holes  with hyperscaling violation. For all our solutions we compute the thermoelectric DC transport coefficients of the corresponding dual field theory. We find that the Solv and Nil black holes without hyperscaling violation are dual to metals while those with hyperscaling violation are dual to insulators.  

\end{abstract}

\tableofcontents

\newpage

\section{Introduction}

Both string theory and condensed matter systems have a large landscape of vacua and we can 
 relate these two using the AdS/CFT correspondence. 
Inspired by this observation the authors of \cite{Iizuka:2012iv} constructed geometries dual to field theories that preserve a generalized notion of translational invariance, following Bianchi classification. In such space-times you can get one point from any other point using an isometry but  contrary to usual translation symmetries the generators of the symmetry do not commute. Instead,  they form a Lie algebra.


In the context of AdS/CMT correspondence one might be interested in studying the behavior of systems at finite temperature. This means to study black hole geometries in the dual gravity theory.
In five dimensions, the event horizon could be a priori any compact orientable 3 dimensional Riemannian manifold.
Nevertheless, due to the Thurston geometrization
conjecture, the event horizon can be
endowed with a metric which is locally isometric to one
of the eight Thurston geometries \cite{Groen:2007zz}. The simplest ones
are given by the Euclidean space $E_3$, the three-sphere $S_3$  the hyperbolic space $H_3$ 
, the products $S_1\times H_2$ and $S_1 \times S_2$. In addition,
there are three non-trivial homogeneous geometries
which are neither constant curvature nor a product
of constant curvature manifolds called the Nil geometry,
the Solv geometry and the geometry of the universal
cover of $SL_2({ \cal R})$ with the following representative metrics
\bea
Solv: \ \ \ \ \ \ d\bar{s}^2&=& e^{2z}dx^2+ e^{-2z}dy^2+dz^2 \,, \nn \\
Nil: \ \ \ \ \ \ d\bar{s}^2&=& dx^2+ dy^2+(dz-xdy)^2 \,, \nn \\
SL_2({ \cal R}): \ \ \ \ \ \ d\bar{s}^2&=& \frac1{x^2}( dx^2+ dy^2 )+ \left(dz+\frac{dy}{x}  \right)^2\,.  
\eea

Interestingly, black holes with Solv and Nil horizons were found to be solutions to  general relativity in the presence of a negative cosmological constant \cite{Cadeau:2000tj}, if one allows the boundary to scale anisotropically in the diferent space-time directions  \cite{Kachru:2008yh,Taylor:2015glc}.
Further solutions with nilgeometry horizons were found in \cite{Hassaine:2015ifa}, where hyperscaling violation asymptotics were studied. In this paper we will further elaborate on these works by considering charged geometries. This is of particular interest in the context of AdS/CMT, since corresponds to studying the dual field theory at finite chemical potential.

In the present paper we will consider the Einstein-Maxwell action in five dimensions
\be 
\tilde S=\int d^5 x \sqrt{-g}\left( \frac{1}{2\kappa^2}\left(   R -2 \Lambda\right) - \frac14 F_{\mu \nu}F^{\mu\nu}\right)+\tilde S_{bdy}\,,
 \label{EM}
\ee
where $\tilde S_{bdy}$ corresponds to some boundary action and we set $\kappa^2=8\pi G_{N}=1$. 
The equations of motion read
\bea
D_\mu F^{\mu\nu}&=&0\,,\nn \\ 
R_{\mu \nu}- \frac12 R\ g_{\mu\nu}+\Lambda \  g_{\mu\nu} & =& -\frac14 F_{\alpha \beta}F^{\alpha\beta} g_{\mu\nu} +  F_{\mu \alpha}F^\alpha_{\,\,\, \nu}\,,
\label{EMeoms}
\eea
and we will look for solutions with charged Solv and Nil horizons that will be dual to field theories at finite temperature $T$ and chemical potential $\mu$ without translational invariance. So we will generalize the solutions of \cite{Cadeau:2000tj,Hassaine:2015ifa} to the case of charged black branes.
Furthermore, we will compute the DC transport properties of the dual field theory giving a first approach to the transport properties of theories with this kind of geometrical duals.

Studying transport coefficients one can do a classification on the different condensed matter theories. A simple example is for instance to study how the DC conductivity $\sigma$ behaves at low temperatures $T\ll \mu$. For metals $\sigma$ decreases as we increase $T$, while the opposite happens for insulators.



Generically, translational invariant systems at a finite chemical potential will have an infinite DC conductivity.
In holography, in order to have a finite conductivity at low temperatures we need to have a geometry that breaks translational invariance or has some other momentum dissipation mechanism. 
A first approach in this direction was done in \cite{Horowitz:2012ky,Horowitz:2012gs,Horowitz:2013jaa}, where operators that depend explicitly on one or more of the spatial coordinates were turned on. This usually leads to solving complicated PDEs. 

Fortunately, there are also clever ways of finding systems with momentum disipation where only ODEs are involved. One may use an internal symmetry as in the Q-lattices \cite{Donos:2013eha,Donos:2014uba}, or by introducing momentum relaxation in an effective way by studying solutions to massive gravity \cite{Vegh:2013sk,Blake:2013bqa,Blake:2013owa,Amoretti:2014zha,Amoretti:2014mma}. A third posibility is to use a Bianchi symmetry \cite{Donos:2012js}. We will follow this path by studying Bianchi $VI_{-1}$ (Solv) and Bianchi $II$ (Nil) symmetries. A remarkable aspect of our model is that we solve the Einstein-Maxwell Lagrangian with a simple ansatz that leads to ODEs and that captures the interesting phenomena of momentum dissipating physics. 


This paper organizes as follows. In Section \ref{sec1} we study charged solvgeometry black holes. Since the solutions are usually numerical, we show first in Section \ref{sec11} some particular analytical dyonic solutions we find by fine-tunning the black holes charge and magnetic field. After this warm up solutions, we construct the numerical charged solutions at zero magnetic field.
In Sections \ref{sec2} and \ref{sec3} we study charged nilgeometry black holes without and with hyperscaling violating exponent.
Furthermore we will compute the thermoelectric transport porperties from horizon data following  \cite{Blake:2013bqa,Donos:2014cya}.

\section{Charged and dyonic solvgeometry black holes }
\label{sec1}
In this section we will study charged and dyonic geometries with Solv horizons.

\subsection{Dyonic solvgeometry black holes}
\label{sec11}
Generically charged and dyonic solvgeometry black holes will require to numerically integrate the Einstein-Maxwell equations of motion (\ref{EMeoms}). Nevertheless, one can tune the magnetic and electric field in order to get analytical solutions.

We will consider the following ansatz for the metric and gauge field
\bea
A&=&A_t(r)\  dt+ A_y\  x\  dy	\, ,  \label{ansatzdy}\nn\\ 
ds^2&=&- r^2 F(r) dt^2 + \frac1{r^2 F(r)} dr^2 + e^{2 z }r^2 dx^2 + e^{-2z}r^2 dy^2+ a_3 dz^ 2\,,
\eea
which corresponds to a dyonic solvgeometry black hole
with anisotropic asymptotic scaling 
\bea
t \rightarrow \lambda\, t\,,\,\,\,\, r \rightarrow \lambda^{-1}\, r \,, \,\,\,\, x\rightarrow \lambda\, x\, , \,\,\,\, y\rightarrow \lambda\, y \,, \,\,\,\, z\rightarrow z \,.
\eea

Non relativistic geometries are interesting in its own right in the context of AdS/CFT because they offer a playground to study dual non-conformal 
field theories \cite{Taylor:2015glc}. 

We find the following solutions
\bea
A_t(r)&=&   \mu \left( 1- \frac{r_h}{r} \right) \,, \nn\\
F(r)&=& 1- \frac{r_h(\mu^2  + r_h^2)}{r^3}+\frac{\mu^2 r_h^4}{r^4}\,,
\eea
where $r_h$ corresponds to the position of the horizon and $\mu$ to the chemical potential of the dual field theory.
The equation of motion for the Maxwell fields is automatically satisfied for $A_y$, nevertheless, consistency of the Einsteins equations requires $A_y= r_h \mu$. This means that charged solvgeometry black holes must be dyonic within the simple metric ansatz  (\ref{ansatzdy}). 
We have also set $\Lambda = -\frac92$ and $a_3=\frac23$.

The back hole temperature and entropy read
\bea
T&=& \frac{3r_h^2-\mu^2}{2\pi r_h}\,,  \nn \\
S&=&2 \pi A_{h}=2 \pi \sqrt{\frac23}r_h^2\, .
\eea
In the $T=0$ limit, the near horizon geometry approaches to $AdS_2\times Solv$.

To finish this section, we would like to make a comment on the electromagnetic duality. If we fix  the coordinate $z=z_0$ to a constant the solution is self dual in the sense
\bea
\star F=F\,,
\label{duality}
\eea
where $\star$ is the  Hodge operator given the four dimensional metric defined by $z=z_0$ \cite{Bravo-Gaete:2017nkp}. 

\subsection{Charged solvgeometry black holes}
\label{sec12}

In order to find charged black hole solutions to the Einstein-Maxwell theory (\ref{EM}) we will have to relax the ansantz  (\ref{ansatzdy}) for the metric. This will lead into more complicated equations of motion and we will have to solve it numerically. Let us then consider the ansatz 
\bea
A&=&A_t(r)\  dt\, ,   \nn\\
ds^2&=&- r^2 N^2(r)F(r) dt^2 + \frac1{r^2 F(r)} dr^2 +r^2 H^2(r)\left( e^{2 z }dx^2 + e^{-2z} dy^2\right)+ \frac{2}{3 H^4(r)} dz^ 2\,.
\eea

The equations of motion read
\bea
\nn
H''&=&\frac{3 r H H' \left(A_t'^2+N^2 \left(-2 F+3 H^4-9\right)\right)+H^2 \left( A_t'^2+9 \left(H^4-1\right)
   N^2\right)+6 r^2 F N^2 H'^2}{6 r^2 F H N^2},\\ \nn
F'&=&\frac{-H^2 A_t'^2-6 r^2 F N^2 H'^2-4 r F H N^2 H'-6 F H^2 N^2-3 H^6 N^2+9 H^2 N^2}{2 r H^2 N^2},\\ 
N'&=&\frac{N H' \left(3 r H'+2 H\right)}{H^2},~ ~ ~~~~~~
A_0''=A_t' \left(\frac{H' \left(3 r H'+2 H\right)}{H^2}-\frac{2}{r}\right).
\eea

For clarity we did not write the $r$ dependence of the functions. Their expansion near the horizon (IR) reads
\bea
H(r)&\approx&h_0-\frac{h_0 \left(a_{t_1}^2+9 \left(h_0^4-1\right) n_0^2\right)}{3 \left(a_{t_1}^2+3 \left(h_0^4-3\right) n_0\right)r_h}(r-r_h)+\ldots, \nn\\ \nn
F(r)&\approx&-\frac{a_{t_1}^2+3 \left(h_0^4-3\right) n_0^2}{2 n_0^2 r_h}(r-r_h)+\ldots,\\ \nn
N(r)&\approx&n_0-\frac{n_0 \left(a_{t_1}^2+9 \left(h_0^4-1\right)n_0^2\right) \left(a_{t_1}^2-3 \left(h_0^4+3\right) n_0^2\right)}{3
   \left(a_{t_1}^2+3 \left(h_0^4-3\right)n_0^2\right)^2r_h}(r-r_h)+\ldots,\\ 
A_t(r)&\approx& a_{t_1}(r-r_h)-\frac{a_{t_1} \left(7 a_{t_1}^4+42 a_{t_1}^2 \left(h_0^4-3\right) n_0^2+27 \left(h_0^8-14 h_0^4+21\right) n_0^4\right)}{6
   \left(a_{t_1}^2+3 \left(h_0^4-3\right)n_0^2\right)^2r_h}(r-r_h)^2+\ldots ,\label{IRSolv}
\eea
while near the boundary (UV) we impose
\bea
H(r)&\approx&1+\ldots\,,\nn\\ \nn
F(r)&\approx&1-\frac{M}{r^3} +\ldots\,,\\ \nn
N(r)&\approx&1+\ldots\,,\\ 
A_t(r)&\approx&\mu-\frac{\rho}{r}+\ldots\,.
\label{bdysolv}
\eea

\begin{figure}[!h]
\begin{center}  
\includegraphics[scale=0.59]{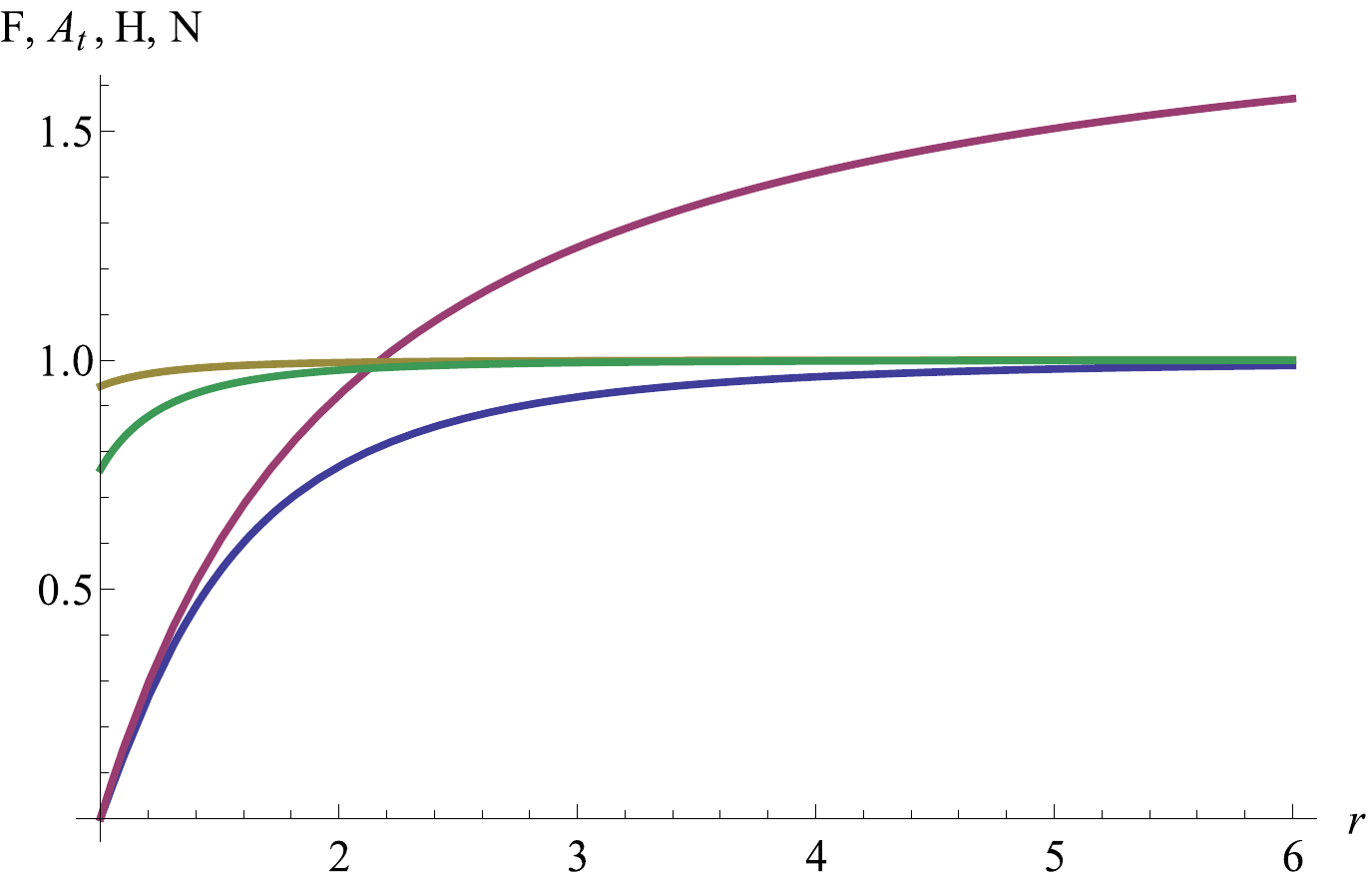}\hfill\includegraphics[scale=0.59]{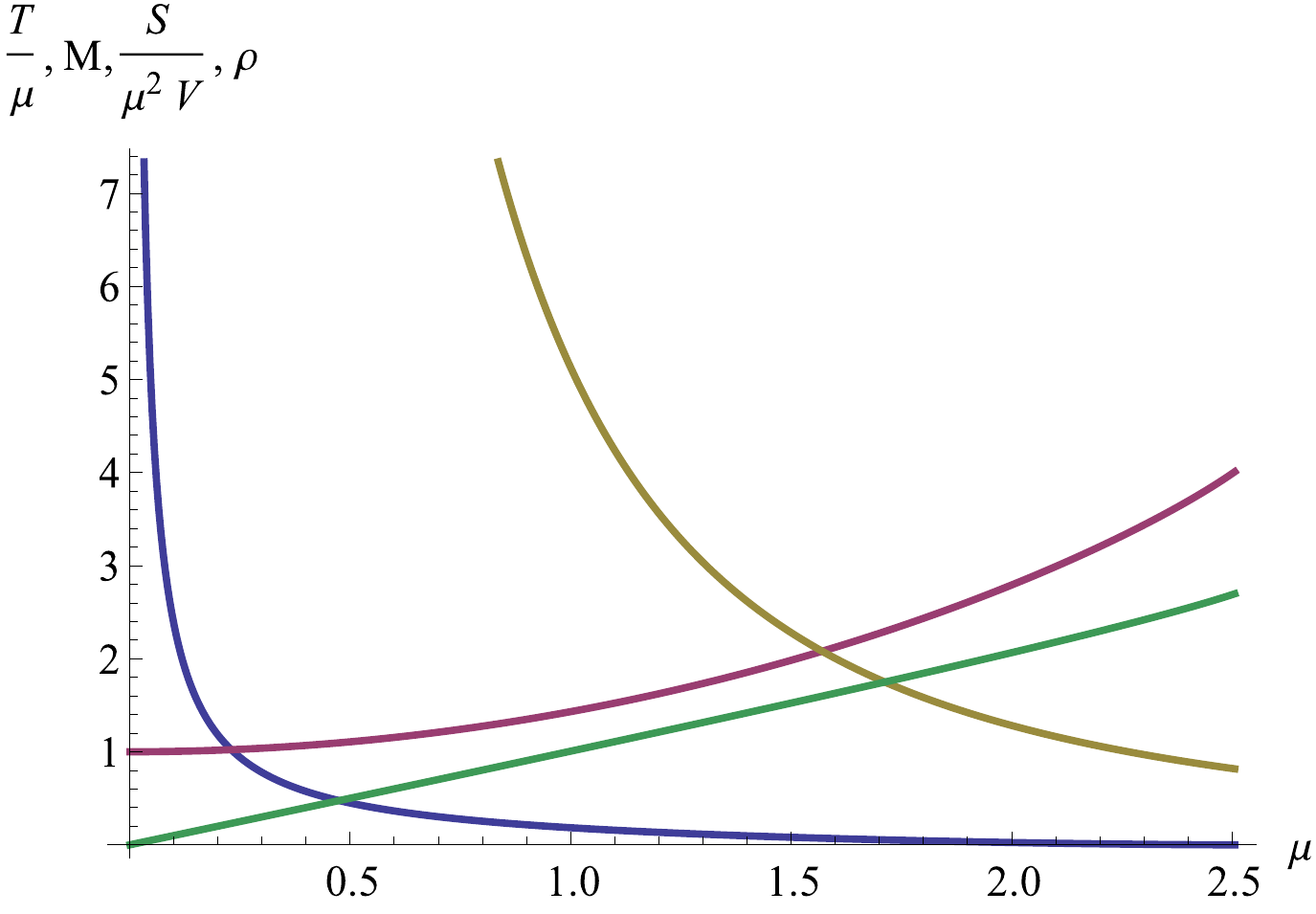}
\caption{Left: Radial profiles for the fields $F$ (Blue), $A_t$ (Purple), $H$ 
(Yellow) and $N$ (Green) for $\mu=1.8$. Right: Termperature $T$ (Blue), black hole mass $M$ (Purple), black hole entropy $S$ (Yellow) and  charge density $\rho$ (Green) for the dual field theory as a function of the chemical potential $\mu$.  \label{fig1}}
\end{center}  
\end{figure}

In the left panel of  Figure \ref{fig1} we show the profiles resulting from integrating the equations of motion. We use a shooting method with the horizon variables $h_0,\,n_0,\, a_{t_1}$  in order to get the desired boundary behavior (\ref{bdysolv}).  For the numerics we set $r_h=1$ without loss of generality. 

The temperature and entropy read
\bea
T&=&-\frac{r_h}{8\pi n_0}\left(a_{t_1}^2+3 \left(h_0^4-3\right) n_0^2\right), \nn\\ 
S&=&2 \pi \sqrt{\frac23}r_h^2
\eea
On the right hand side of  Figure \ref{fig1}, we show the dependence of the black hole's temperature, mass and entropy as a function of the chemical potential of the dual field theory. We plot the dual's field theory charge density. Along this paper we will usually consider the dimensionless ratio $T/\mu$, to plot physical quantities of the dual field theory.

\subsection{Finite conductivities from charged solvgeometry black holes}
\label{sec13}

Consider a system at equilibrium at finite chemical potential and temperature.
The addition of a small electric field $E_i$ or thermal gradient $\nabla_i T$ will induce an electric current $J ^i$ and a heat current $Q^i=T^{ti}-\mu J^i$, where $T^{ij}$ is the stress tensor of the dual field theory. At linearized order, the response is controlled by the Ohm/Fourier law 
\bea
\left(\begin{matrix}
 J \\
 Q
\end{matrix}\right)=\left(\begin{matrix}
   \sigma    & \alpha T \\
  \bar\alpha T    &\bar\kappa T
\end{matrix}\right) \left(\begin{matrix}
   E \\
- \nabla T/T
\end{matrix}\right)\,,
\label{ohm}
\eea
defining the electric conductivity $\sigma$, the thermoelectric conductivities $\alpha$, $\bar\alpha$ and the thermal conductivity $\bar\kappa$.

Systems with translation invariance and finite charge density have an infinite $DC$ conductivity. Nonetheless, in the directions where the translation invariance is broken, we expect a finite $DC$  conductivity. That will be the $z$ direction in our solvgeometry charged black holes or the $x$ direction in our nilgeometry charged black holes. We will read then the coefficients of the matrix (\ref{ohm})  from horizon data, following the method developed in \cite{Donos:2014cya}.

The holographic dictionary gives us the expressions for the electric and heat current in the dual field theory \cite{Liu:2017kml,Donos:2017oym}
\bea
J&=&\sqrt{-g} F^{ri}\,,\nn\\ 
Q&=&\sqrt{-g}G^{ri} + J A_t,\label{JQ}
\eea
where the tensor $G^{\mu\nu}$ reads
\be 
G^{\mu\nu}=\nabla^\mu k^\nu +\frac{1}{3}k^{[\mu}F^{\nu]\sigma}A_\sigma, 
\ee
where $k=\partial_t$ and the index $i$ denotes the direction on which the electric field is applied. It will be the $z$ direction for the solvegeometry and $x$ for the Nil cases. 

Now we proceed to compute these transport coefficients for the system dual to the geometries studied in Section \ref{sec12}.

\subsubsection{Calculating $\sigma$ and $\bar \alpha$}

In order to compute $\sigma$ and $\bar \alpha$ we need to study linear response of the black hole after small perturbations of the background and gauge field,
\bea
\delta A&=&(-Et+\delta a_{z}(r)) dz\,, \nn \\ 
\delta ds^2&=& \  2\delta g_{tz}(r)\  dt \, dz+ 2\delta h_{rz}(r) \ dr \,dz\,,
\eea
the constant $E$ parametrizes an applied (DC) electric field. With this ansatz for the fluctuations we have to solve two non trivial Maxwell equations. From one of them we can build the explicit expression for the electric current $J$ \eqref{JQ}
\bea
J= \frac{\sqrt{\frac{3}{2}} r^2 H^4 \left(\delta g_{tz} A_t'+r^2 F N^2 \delta a_z'\right)}{N}\, .
\eea
We check that the remaining Maxwell equation for $\delta a_z$ is equivalent to $\partial_r J=0$. Then we can evaluate $J$ at any $r$, in particular at the black hole horizon. Moreover the heat current \eqref{JQ} is
\be
Q=J A_t+\sqrt{\frac{3}{8}}r^6F^2N^3H^4\left(\frac{\delta g_{tz}}{r^2N^2F}\right)'.
\ee

The Einstein equations for the fluctuations are
\bea
\delta h_{rz}&=&-\frac{E A_t'}{3 r^2 F H^4 N^2}\,, \nn\\
Q' &=&0\,.
\eea
For a free falling observer the horizon of a black hole is a regular place and then the electromagnetic field must be regular. 
This means that $A_\mu$ must depend of $r$ and $t$ through the non-singular combination $dv=dt+\sqrt{\frac{g_{rr}}{g_{tt}}}dr$. Near the horizon then the Eddington-Finkelstein coordinate is
\be
v\sim t+\frac{1}{r_h^2n_0 f_1}\log (r-r_h)\,,\label{EddingSol}
\ee
with $f_1$ and $n_0$ the coefficients in the expansion of the function $F$ and $N$ in the IR \eqref{IRSolv}. The equation \eqref{EddingSol} is going to be valid for the Nil geometries that we will study in the next section too. This allows us to fix the value of $a_z'(r_h)$,  $\delta g_{tz}(r_h)$  and   $\delta g_{tz}'(r_h)$, fixing the value of $J$ and $Q$  in terms of the background fields at the horizon 
\bea
J&=&\frac{E \left(a_{t_1}^2+3 h_0^4 n_0^3 r_h^2\right)}{\sqrt{6} n_0^3}\,, \nn\\
Q&=&\frac{a_{t_1} r_h E  \left(a_{t_1}^2+3 \left(h_0^4-3\right) n_0^2\right)}{4 \sqrt{6} n_0^3} \,.
\eea
We can now read the electric and thermal conductivities

\bea
\sigma&=&\frac{\partial J}{\partial E}=\frac{a_{t_1}^2+3 h_0^4 n_0^3 r_h^2}{\sqrt{6} n_0^3},\nn\\
\bar\alpha&=&\frac{1}{T}\frac{\partial Q}{\partial E}=\frac{\sqrt{\frac{2}{3}} \pi a_{t_1}}{n_0^2}.
\eea

\begin{figure}[!h]
\begin{center}  
\includegraphics[scale=0.65]{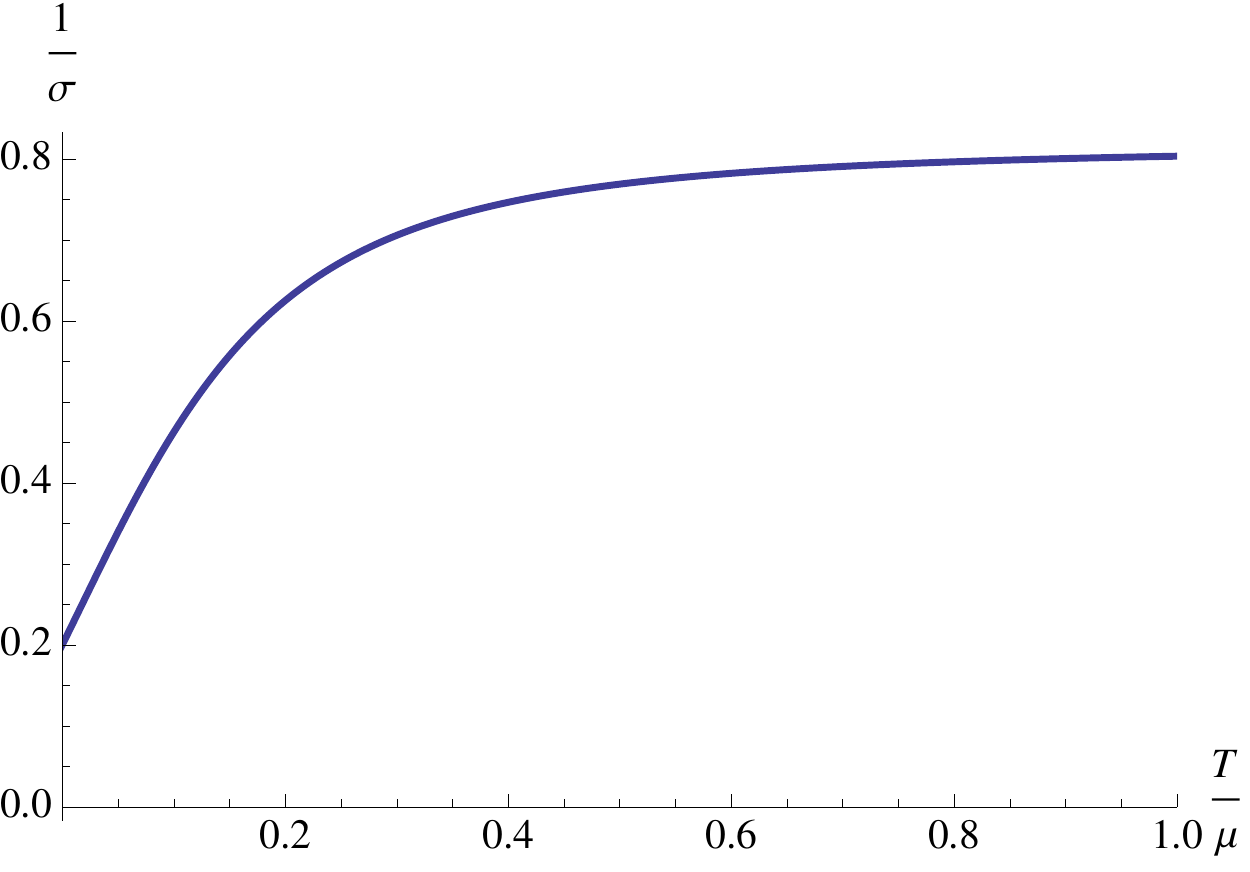}\hfill \includegraphics[scale=0.65]{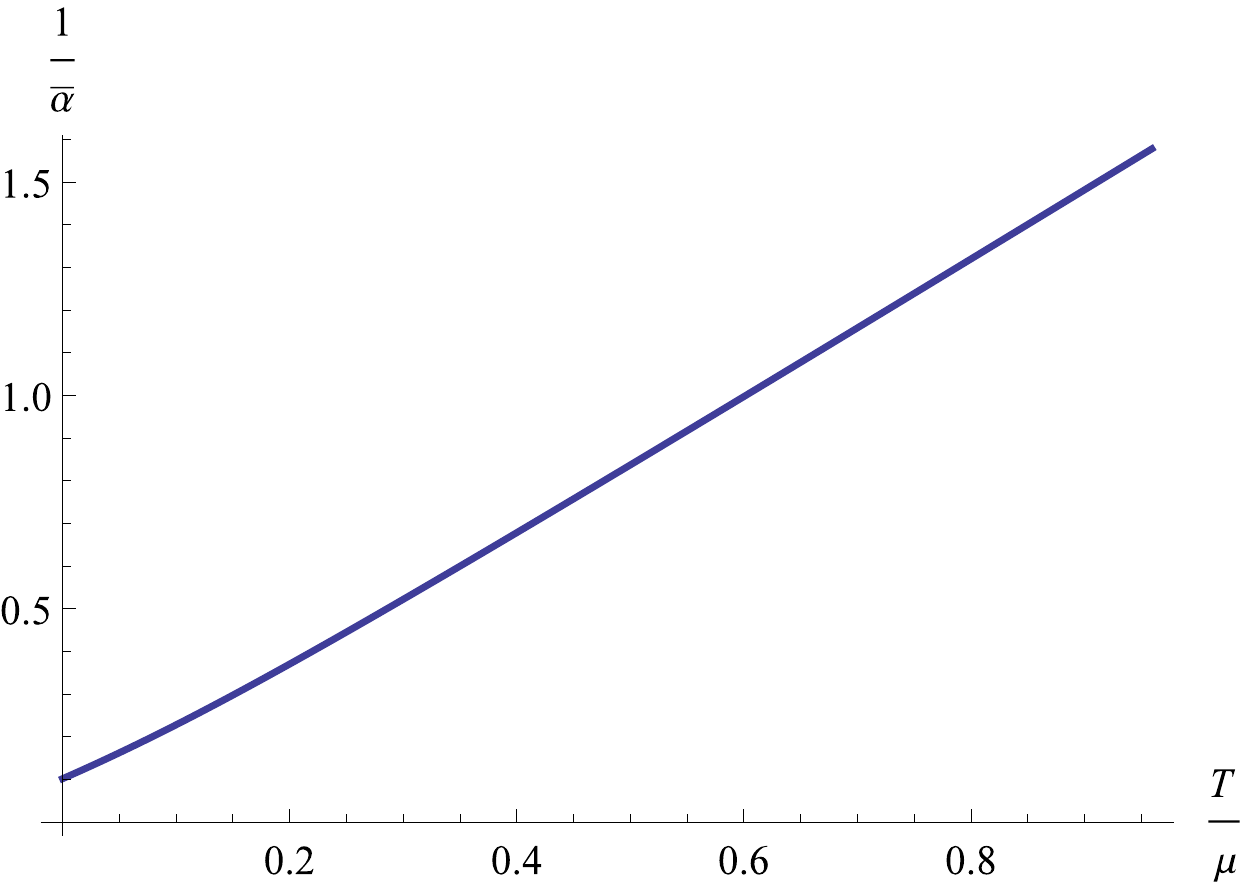}
\caption{Resistivites $1/\sigma$ and $1/\bar \alpha$ as a function of the temperature.\label{fig1b}}
\end{center}  
\end{figure}

In Figure \ref{fig1b} we plot the resistivities $1/\sigma$ and $1/\bar\alpha$ as a function of the temperature. In order to do this, we use the values for the background horizon variables $h_0,\,n_0,\, a_{t_1}$ that give us the desired boundary behavior (\ref{bdysolv}).

We observe that $1/\sigma \sim 0.2 +  T$ at low $T$ and saturates to a constant at hight $T$. This increasing with the temperature behavior for the electrical resistivity is typical of a metal. We can interpret the fact that the thermal resistivity saturates for high temperatures as a strongly interacting version of the Mott-Ioffe-Regel  bound \cite{Gunnarsson:2003zz}, though the saturation value might not be the expected for weakly interacting systems. 

For the thermal resistivity  $1/\bar\alpha$ we see that it goes to a constant at low $T$ while it grows linearly for high $T$. This will be a common feature for all the holographic systems we are going to study along this paper.

Further interesting phenomenology will arise from the study of  $\alpha$ and $\bar \kappa$, which we will proceed to compute in the next subsection.

\subsubsection{Calculating $\alpha$ and $\bar \kappa$}

The fluctuations we are interested in now read\footnote{ It is interesting to observe that after theese perturbations are turned on, the vector $k=\partial_t$ is no longer a Killing vector and $\nabla_\mu k^\mu\neq 0$. Therefore we cannot extend straightfowardly the results of \cite{Banks} to obtain the transport coefficients in a general way from the background geometry and fields.}
\bea
\delta ds^2&=&2(t \delta f_2(r)+\delta g_{tz})dt dz+ 2\delta h_{rz}drdz\,,\nn \\
\delta A&=&(-t \delta f_1(r)+\delta a_z(r))dz\,.
\eea

Following the arguments of the previous subsecition we will study the following objects that do not depend on the radial coordinate
\bea
J&=&\frac{\sqrt{\frac{3}{2}} r^2 H^4 \left(A_t' (\delta g_{tz}+t \delta f_2)+r^2 F N^2 \left(\delta a_z'-t \delta
f_1'\right)\right)}{N}\,,\nn\\
Q&=&J A_t+\sqrt{\frac{3}{8}}F^2N^3r^6H^4\left[t\left(\frac{\delta f_2}{N^2F r^2}\right)'+\left(\frac{\delta g_{tz}}{N^2F r^2}\right)'\right]\,.
\eea

We can erase their temporal dependence choosing
\bea
\delta f_1&=&E+\zeta A_t \,, \nn\\
\delta f_2&=& \zeta r^2 N^2 F.
\eea
The remaining Einstein equation reads
\be 
\delta h_{rz}=\frac{H \left(\delta f_2'-2 \delta f_1 A_t'\right)+4 \delta f_2 H'}{6 r^2 F H^5 N^2}\,.
\ee
Because $J$ and $Q$ are constants we can evaluate it on the horizon. In order to do that we use the Eddington-Filkenstein coordinates \eqref{EddingSol} to obtain the near horizon behavior of $\delta g_{tz}$ and $\delta a_z$
\bea
\delta a_z&\sim & \frac{E}{r_h^{2}n_0^2 f_1}\log(r-r_h)\,, \nn \\
\delta g_{tz}&\sim & -\left(\delta g_{rz}\right)_{_{r\rightarrow r_h}}-\zeta\frac{r^2 N^2F}{r_h^2  n_0 f_1}\log(r-r_h)\,.
\eea
Using this we have
\bea
J&=&\frac{4 E \left(a_{t_1}^2+3 h_0^4 n_0^3 r_h^2\right)+a_{t_1} \zeta   \left(a_{t_1}^2+3 \left(h_0^4-3\right)
   n_0^2\right)r_h}{4 \sqrt{6} n_0^3}\,, \nn \\
   Q&=&\frac{r_h \left(a_{t_1}^2+3 \left(h_0^4-3\right) n_0^2\right) \left(\zeta  r_h \left(a_{t_1}^2+3 \left(h_0^4-3\right)
   n_0^2\right)+4 a_{t_1} E \right)}{16 \sqrt{6} n_0^3},
\eea
and from here we can straightfowardly obtain the transport coefficients
\bea
\alpha&=&\frac{1}{T}\frac{\partial J}{\partial\zeta}=\frac{\sqrt{\frac{2}{3}} \pi a_{t_1}}{n_0^2},\nn\\
\bar\kappa&=&\frac{1}{T}\frac{\partial Q}{\partial\zeta}=-\frac{\pi  r_h \left(a_{t_1}^2+3 \left(h_0^4-3\right) n_0^2\right)}{2 \sqrt{6} n_0^2},
\eea
expressed in terms of the background fields in the horizon.
A nice check is to observe that after a cumbersome computation we find $\alpha=\bar\alpha$, as it should  be since the conductivities matrix (\ref{ohm}) is symmetric.

Another interesting quantity is the thermal conductivity at zero electric current
\bea
\kappa=\bar\kappa-\frac{\alpha \bar\alpha T}{\sigma}=-\frac{\sqrt{\frac{3}{2}} \pi  h_0^4 n_0 r_h^3 \left(a_{t_1}^2+3 \left(h_0^4-3\right) n_0^2\right)}{2 \left(a_{t_1}^2+3
   h_0^4 n_0^3 r_h^2\right)}.
\eea

\begin{figure}[!h]
\begin{center}  
\includegraphics[scale=0.64]{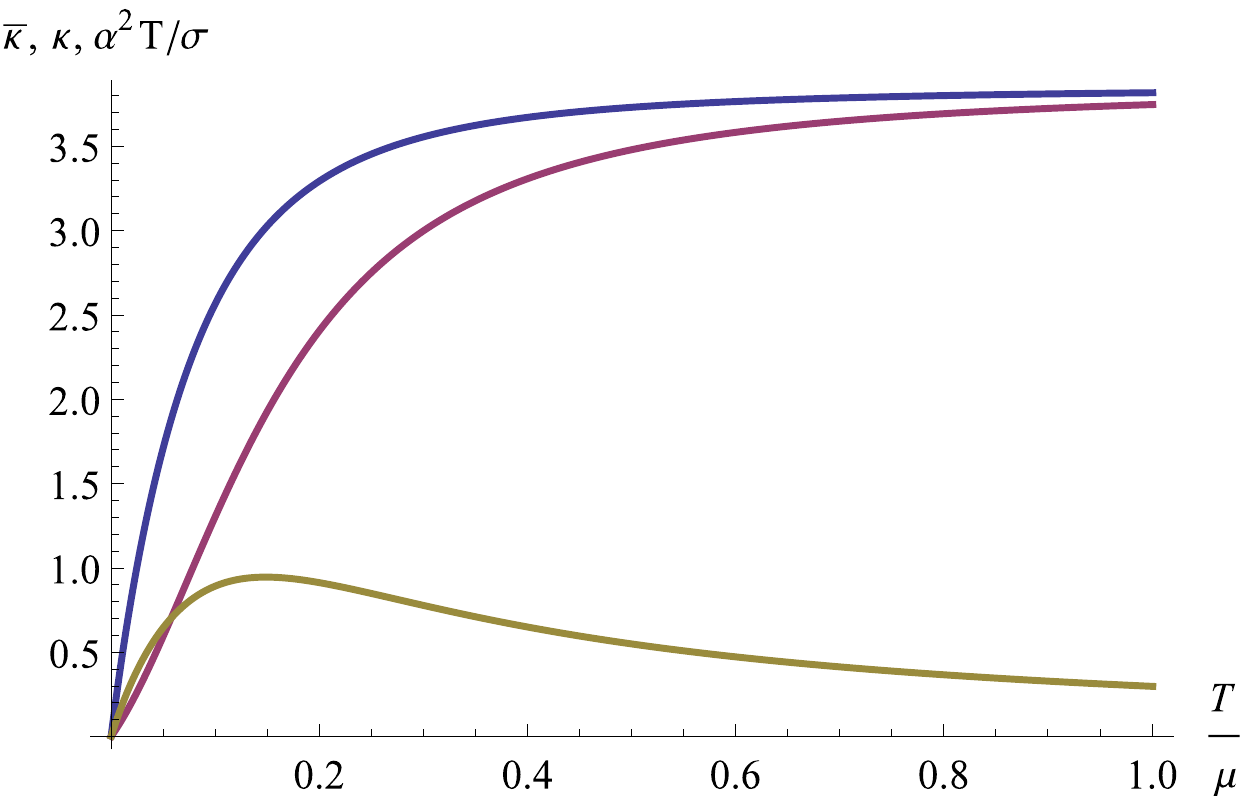}\hfill \includegraphics[scale=0.64]{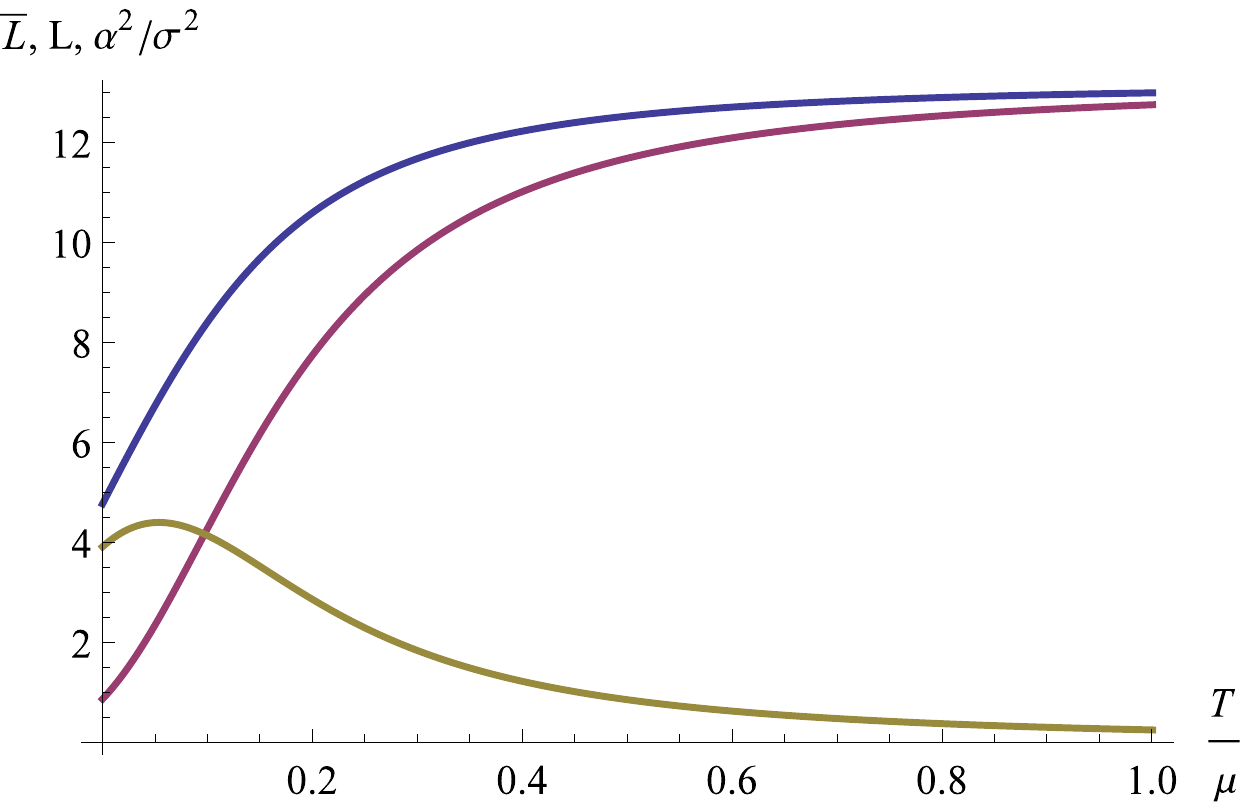}
\caption{Left: Thermal conductivities $ \bar \kappa$ (Blue) and $ \kappa$ (Purple) and the difference between the two of them $\alpha^2 T/\sigma$ (Yellow) as a function of the reduced temperature. Right:  Wiedemann-Franz ratios $\bar L$ (Blue) and $L$ (Purple) and the difference between these two $\alpha^2/\sigma^2$ (Yellow)  as a function of the temperature. \label{fig1c}}
\end{center}  
\end{figure}

Again we can proceed to compute the numerical values for the thermal conductivities $\bar \kappa$ and $\kappa$ which we show in Figure \ref{fig1c}. We find that for low temperatures both  $\bar \kappa$ and $\kappa$ behave linear in $T$,  $\bar \kappa\sim \kappa \sim T$. Following naively  the classification of \cite{Mahajan:2013cja} our low temperature regime satisfying  $\bar \kappa\sim \kappa \sim \alpha^2 T/\sigma$ would correspond, if the charge carriers were fermions, to a non-Fermi liquid with patchwise conserved momenta.
As we increase the temperature the two coefficients become more and more different, so that the difference between the two of them $\frac{\alpha \bar\alpha T}{\sigma}$ develops a maximum for intermediate temperatures. At large $T$ we see that both $\bar \kappa$ and $\kappa$  saturate to the same constant. This  $\kappa\sim \bar \kappa  \gg  \alpha^2 T/\sigma$ is usually associated to long-lived quasi-particles.

Also of particular interest are the ratios
\bea
\bar L&=&\frac{\bar\kappa}{\sigma T}=\frac{4 \pi ^2 n_0^2}{a_{t_1}^2+3 h_0^4 n_0^3 r_h^2},\nn\\
L&=&\frac{\kappa}{\sigma T}=\frac{12 \pi ^2 h_0^4 n_0^5 r_h^2}{\left(a_{t_1}^2+3 h_0^4 n_0^3 r_h^2\right)^2},
\eea
where $L$ gives the Wiedemann-Franz law, and is a constant for systems where charge carriers are the responsible for heat transport. We plot these ratios in the right hand side panel of Figure \ref{fig3b}. We observe that at low temperatures both $L$ and $\bar L$ go to a constant and increase linearly in $T$. More precisely $\bar L\sim 4.76+3.89 T$, while $L \sim 0.86+3.50 T$. As we increase the temperature both $L$ and $\bar L$ both saturate to the same value, indicating that charge carriers are the responsible for heat transport at high temperatures.

We see that the crossover between non-saturating and  saturating behaviors for $L$, $\bar L$, $\kappa$, $\bar \kappa$ and $1/\sigma$ occurs approximately at the same temperature scaler for all these transport coefficients. We might think then that our geometries might be dual to a strongly correlated metal for which the charge carriers behavior interpolates between a non-Fermi liquid with patchwise conserved momenta and  long-lived quasi-particles as we increase the temperature.

\section{Charged nilgeometry black holes}
\label{sec2}

\subsection{Solutions}
\label{sec21}
We will now consider charged black holes solutions with Nil horizon geometry. Let us then consider the ansatz 
\bea
A&=&A_t(r)\  dt\, , \label{ansatznil} \nn\\ 
ds^2&=&- r^{3} N^2(r)F(r) dt^2 + \frac1{r^2 F(r)} dr^2 +r^2 H^2(r)\left( dx^2 +  dy^2\right)+ \frac{11}{2}r^4\left(   dz-x dy   \right)^ 2\,.
\eea
Using $\Lambda = -\frac{99}{8}$ the equations of motion read
\bea
A_t''&=&-\frac{A_t' \left(4 H^4 A_t'^2 \left(H-r H'\right)+3 r N^2 \left(8 r^2 F H^3 H'^2+r (56 F+33) H^4 H'+(56 F-33)
   H^5+11 r H'+33 H\right)\right)}{24 r^2 F H^4 N^2 \left(r H'+2 H\right)}\, .  \nn \\
 H''&=&\frac{3 r N^2 \left(8 r^2 F H^3 H'^2-r (8 F+33) H^4 H'-11 r H'+33 H^5-33 H\right)-4 H^4 A_t'^2 \left(H-r H'\right)}{24 r^3 F H^4 N^2},\nn\\ \nn
F'&=&-\frac{4 H^4 A_t'^2 \left(2 r H'+H\right)+3 r N^2 \left(24 r^2 F H^3 H'^2+2 r (40 F-33) H^4 H'+11 (8 F-3) H^5-22 r H'-55 H\right)}{24 r^2 H^4 N^2 \left(r H'+2 H\right)},\\ 
N'&=&\frac{3 r N^2 \left(8 r^2 F H^3 H'^2+r (4 F-33) H^4 H'-11 r H'+33 H^5-33 H\right)-4 H^4 A_t'^2 \left(H-r H'\right)}{24 r^2 F H^4 N \left(r H'+2 H\right)}\,.
\eea

Near the horizon the functions behave as
\bea
\nn
A_t(r)&\approx&a_{t_1}(r-r_h)-\frac{a_{t_1}\left(80 a_{t_1}^4 h_0^8-1320 a_{t_1}^2 \left(3 h_0^4+1\right)h_0^4 n_0^2r_h+1089 \left(45 h_0^8+6
 h_0^4+5\right) n_0^4r_h^2\right)}{2r_h \left(4 a_{t_1}^2 h_0^4-33 \left(3 h_0^4+1\right) n_0^2r_h\right)^2}(r-r_h)^2+\ldots \, \\
H(r)&\approx&h_0+\frac{4 a_{t_1}^2 h_0^5-99 h_0 \left(h_0^4-1\right) n_0^2r_h}{r_h\left(4 a_{t_1}^2 h_0^4-33 \left(3 h_0^4+1\right)n_0^2r_h\right)}(r-r_h)+\ldots,\nn\\ \nn
F(r)&\approx&\frac{1}{24r_h^2} \left(-\frac{4 a_{t_1}^2}{n_0^2}+\frac{33r_h}{h_0^4}+99r_h\right)(r-r_h)+\ldots,\\ 
N(r)&\approx&n_0+\frac{16 a_{t_1}^4 h_0^8 n_0+264 a_{t_1}^2 \left(7-3 h_0^4\right) h_0^4 n_0^3r_h+3267 \left(3 h_0^8+2 h_0^4-5\right)
  n_0^5r_h^2}{2 r_h\left(4 a_{t_1}^2 h_0^4-33 \left(3 h_0^4+1\right) n_0^2r_h\right)^2}(r-r_h)+\ldots \, . \label{IRNil}
\eea

We will shoot from the horizon towards the boundary looking for solutions that has the following form
\bea
H(r)&\approx&1+\ldots\,,  \nn \\ \nn
F(r)&\approx&1-\frac{M}{r^{11/2}}+\ldots\,,\\ \nn
N(r)&\approx&1+\ldots\,,\\
A_t(r)&\approx&\mu-\frac{\rho}{r^{5/2}}+\ldots\,.
\label{bdynil}
\eea
This sets $\Lambda=-99/8$.
This solutions correspond asymptotically to Nil geometries with anisotropic asymptotic scaling 
\bea
t \rightarrow \lambda^{3/2}\, t\,,\,\,\,\, r \rightarrow \lambda^{-1} r \,, \,\,\,\, x\rightarrow \lambda\, x\, , \,\,\,\, y\rightarrow \lambda\, y \,, \,\,\,\, z\rightarrow \lambda^2\, z \,.
\eea

\begin{figure}[!h]
\begin{center}  
\includegraphics[scale=0.59]{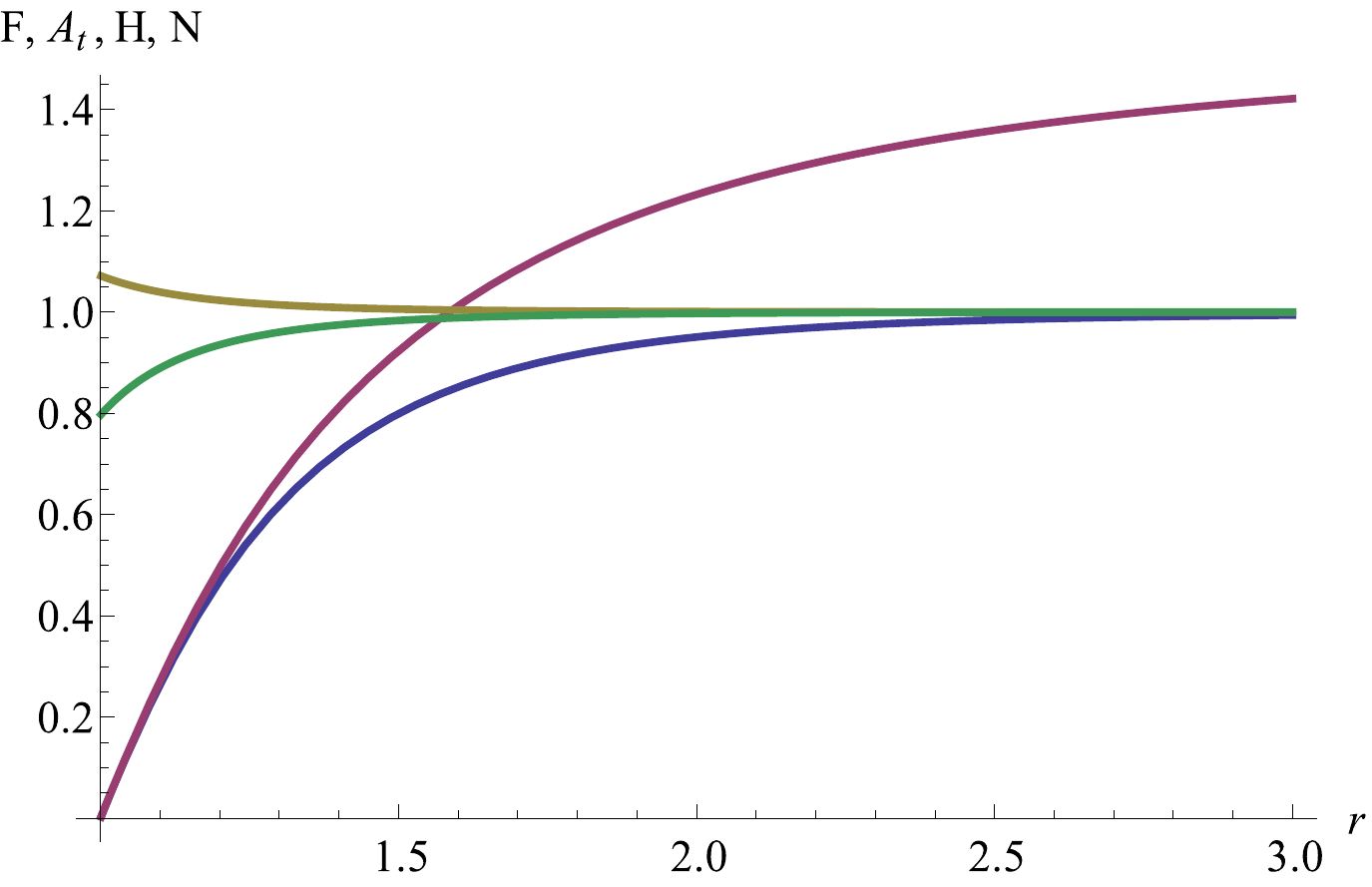}\hfill\includegraphics[scale=0.59]{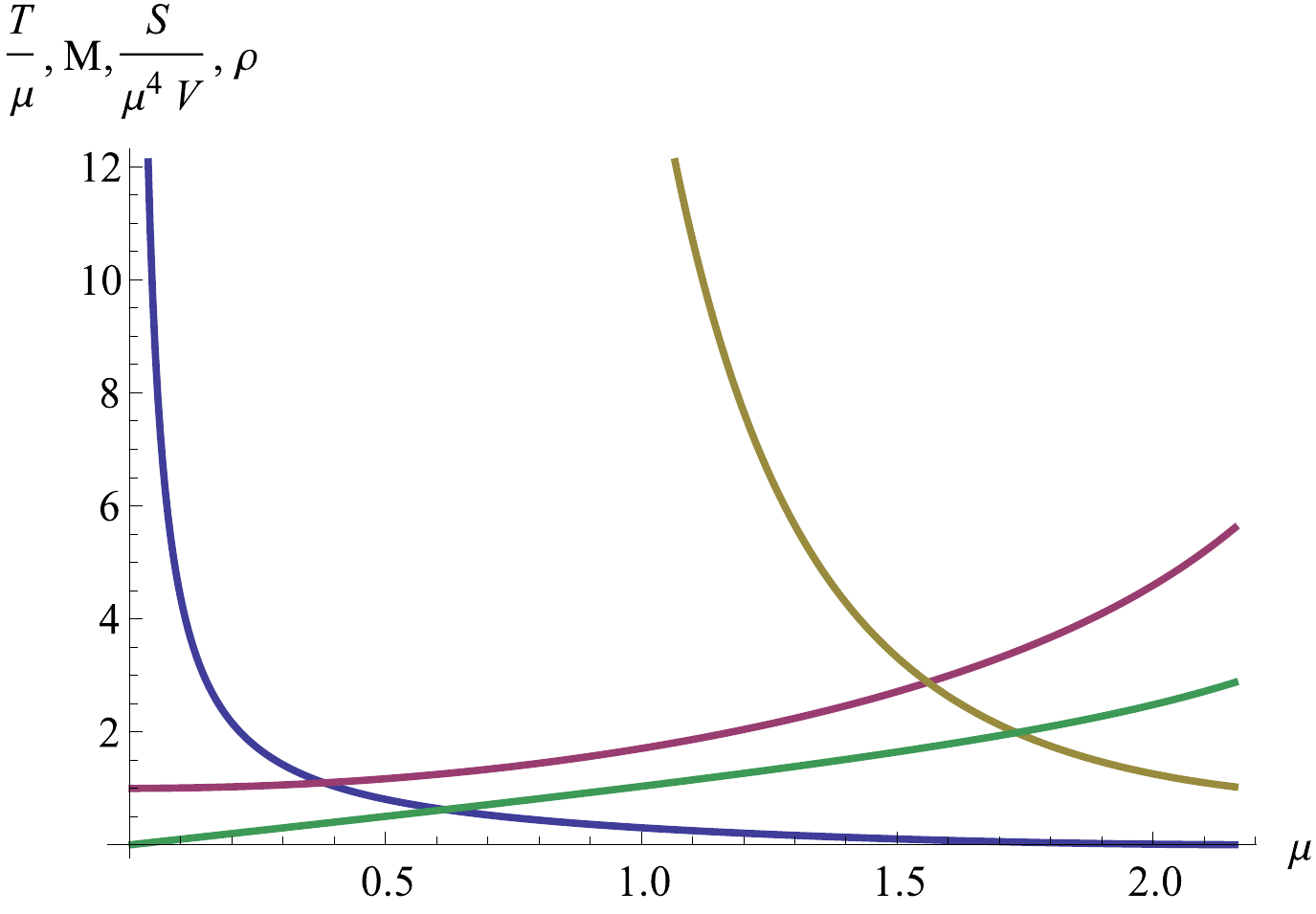}
\caption{Left: Radial profiles for the fields $F$ (Blue), $A_t$ (Purple), $H$ 
(Yellow) and $N$ (Green) for $\mu=1.53$. Right: Termperature $T$ (Blue), black hole mass $M$ (Purple), black hole entropy $S$ (Yellow) and  charge density $\rho$ (Green) for the dual field theory as a function of the chemical potential $\mu$.  \label{fig2}}
\end{center}  
\end{figure}

In the left panel of  Figure \ref{fig2} we show the profiles resulting from integrating the equations of motion. We use a shooting method with the horizon variables $h_0,\,n_0,\, a_{t_1}$  in order to get the desired boundary behavior (\ref{bdynil}).

The temperature and entropy read
\bea
T&=&\frac{r_h}{96\pi } n_0\left(-\frac{4 a_{t_1}^2}{n_0^2}+33 r_h\left(\frac{1}{h_0^4}+3\right)\right), \nn\\
S&=&2\pi\sqrt{\frac{11}{2}}r_h^4h_0^2
\eea
On the right hand side of  Figure \ref{fig2}, we show the dependence of the black hole's temperature, mass and entropy as a function of the chemical potential of the dual field theory. We also plot the dual's field theory charge density.

\subsection{Finite conductivity from charged nilgeometry black holes}
\label{sec21}

\subsubsection{Calculating $\sigma$ and $\bar \alpha$}
In this section we will repeat the procedure realized before in the charged Solvgeometry to obtain the transport coefficients of the dual field theory. 
We start computing $\sigma$ and $\bar \alpha$ and in order to do this we  consider small perturbations around (\ref{ansatznil}) of the form
\bea
\delta A&=&(-Et+\delta a_{x}(r)) dx\,, \nn \\ 
\delta ds^2&=& 2\  \delta g_{tx}(r)\  dt \, dx+ 2 r^2 H^2 \delta h_{rx}(r) \ dr \,dx\,.
\eea
The electric current $J$ reads
\bea
J=\sqrt{\frac{11}2}\frac{r^{3/2}}{N}\left( \delta g_{tx} A_t' + r^3 N^2 F \delta a_x' \right) \, ,
\eea
and  the Maxwell equation for $a_x$ is equivalent to $\partial_r J=0$. This implies that we can evaluate $J$ at any $r$, including at the black hole horizon. 

\begin{figure}[!h]
\begin{center}  
\includegraphics[scale=0.61]{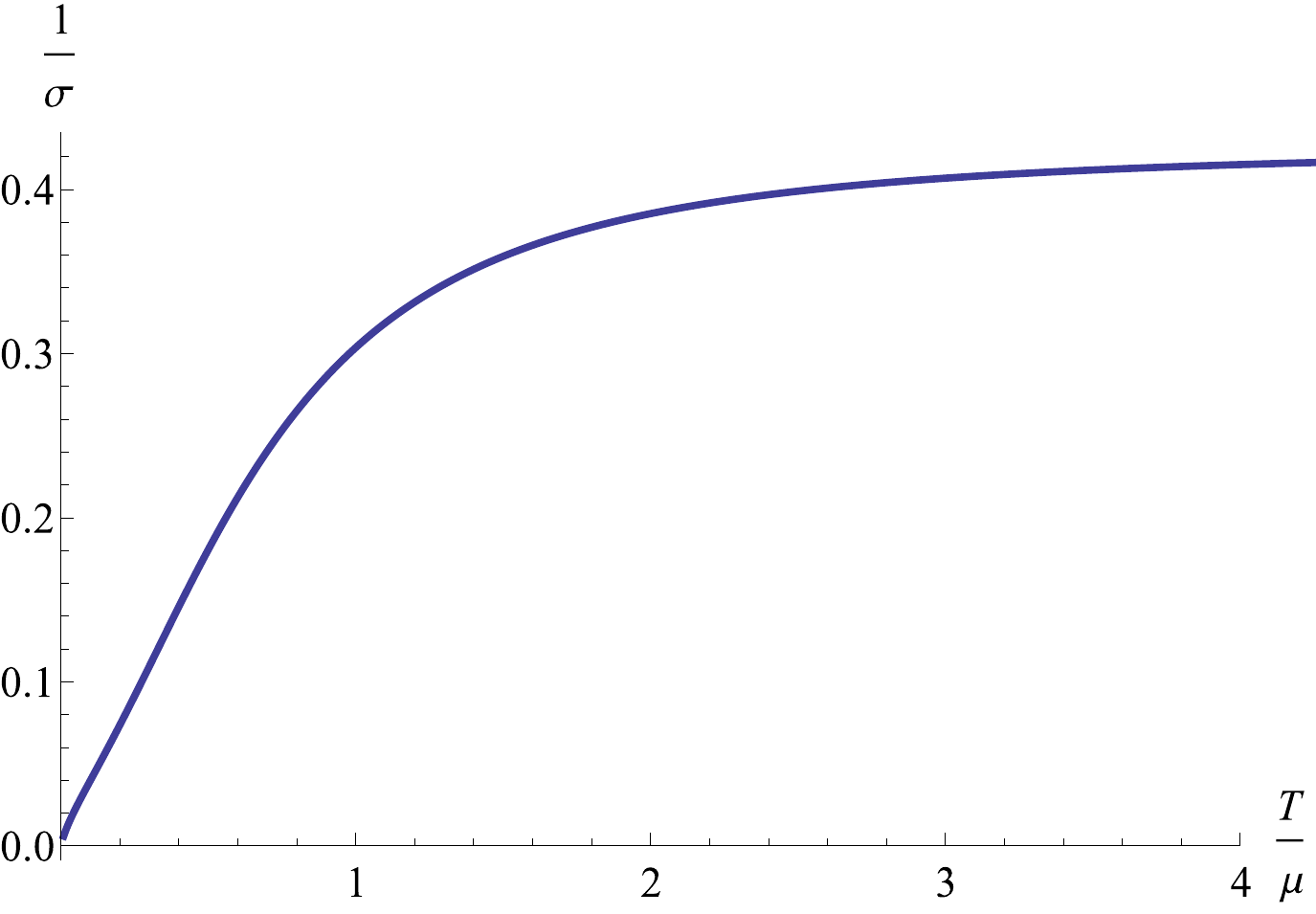}\hfill \includegraphics[scale=0.61]{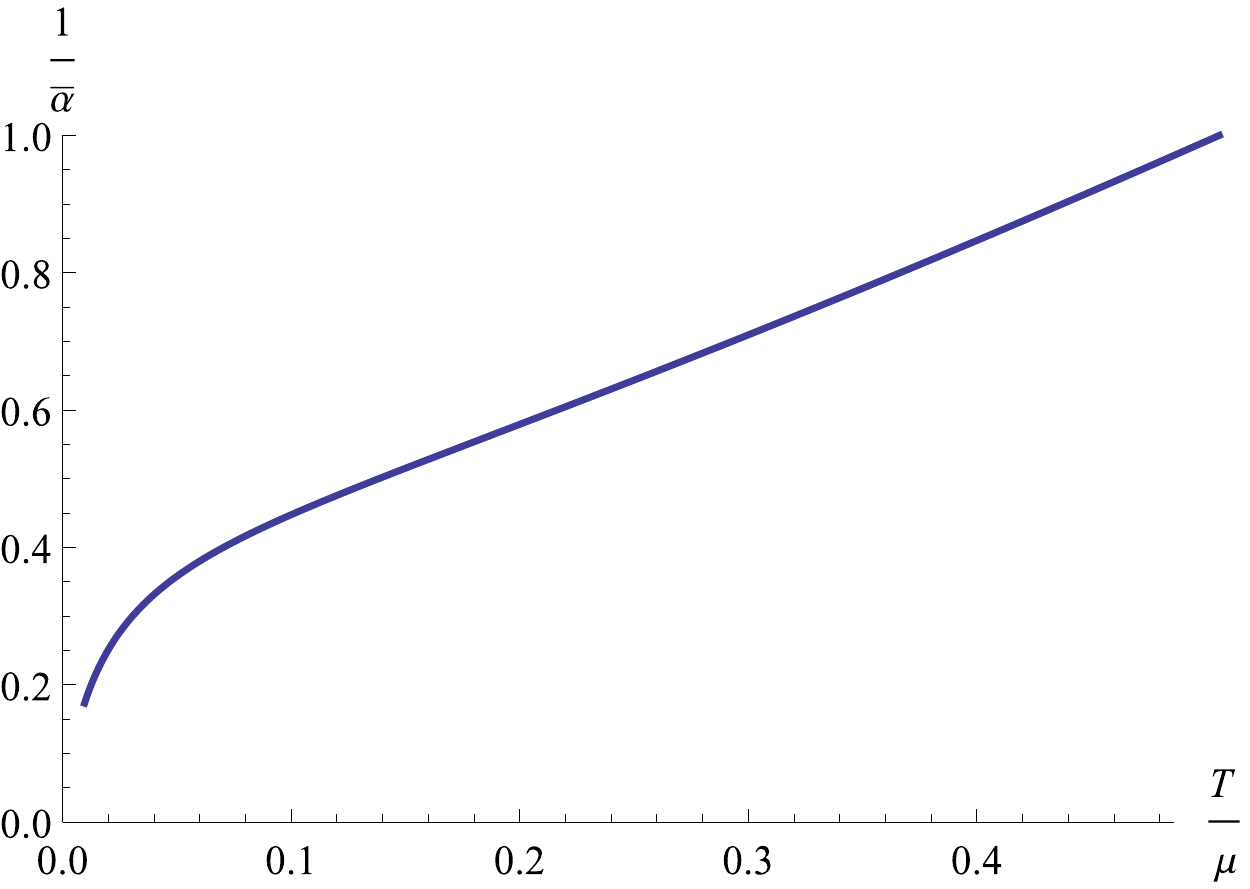}
\caption{Resistivites $1/\sigma$ and $1/\bar\alpha$ as a function of the temperature.\label{fig3}}
\end{center}  
\end{figure}

The heat current $Q$ reads
\be
Q=J A_t+\sqrt{\frac{11}{8}}r^{\frac{15}{2}}F^2N^3\left(\frac{\delta g_{tx}}{r^3N^2F}\right)'\,,
\ee
and the Einstein equations read
\bea
\delta h_{rx}&=&-\frac{4 E H^2 A_t' }{11 r^5 N^2 F}\,, \nn\\
Q' &=&0\,.
\eea
Again, using the Eddington-Finkelstein coordinates $(v,u)$ we have that near the horizon
\be
v\sim t+\frac{1}{r_h^{5/2}n_0 f_1}\log (r-r_h)\label{Edding}
\ee
with $f_1$ the coefficient of the expansion of the function $F$ in the IR \eqref{IRNil}. Asking for regularity of the fluctuations in the near horizon we obtain the values of $a_x'(r_h)$,  $\delta g_{tx}(r_h)$  and   $\delta g_{tx}'(r_h)$ Now we can write  $J$ and $Q$  in terms of the background fields at the horizon 
\bea
J&=& E\, \frac{11 r_h^{7/2}n_0^3+4 a_{t_1}^2 h_0^4 }{\sqrt{22}n_0^3r_h^{3/2}}\,, \nn\\
Q&=& E  \frac{\left(4 a_{t_1}^2 h_0^4-33 \left(3 h_0^4+1\right) n_0^2 r_h\right)a_{t_1}}{12 \sqrt{22 r_h}  n_0^3}\,.
\eea
From these we can obtain the following transport coefficients 
\bea
\sigma &=&\frac{\partial}{\partial E}J=\frac{11 r_h^{7/2}n_0^3+4 a_{t_1}^2 h_0^4 }{\sqrt{22}n_0^3r_h^{3/2}}\,, \nn\\
\bar\alpha &=&\frac{1}{T}\frac{\partial}{\partial E}Q=\frac{4 \sqrt{\frac{2}{11}} a_{t_1} h_0^4 \left(4 a_{t_1}^2 h_0^4-33 \left(3 h_0^4+1\right) n_0^2 r_h\right)}{\pi  n_0^2
  r_h^{3/2} \left(33 \left(3 h_0^4+1\right) n_0^2 r_h-a_{t_1}^2 h_0^4\right)}\,,
\eea
which we plot in Figure \ref{fig3}.

Numerically studying this transport coefficients we find intereting behaviors with the temperature. We find a linear in $T$ electrical resistivity $1/\sigma\sim T$ for low temperatures. On the other hand, $1/\sigma$ saturates to a constant value for large temperatures. This means that our system satisfies a Mott-Ioffe-Regel bound. The thermal resistivity at finite electric field $1/\alpha$ has also a crossover between two different behaviors but at a different temperature, as can be seen from Figure \ref{fig3}. For low temperatures $1/\alpha\sim T^{1/2}$ while for large $T$, $1/\alpha\sim T$.

Further interesting phenomenology will arise from the study of  $\alpha$ and $\bar \kappa$, which we will proceed to compute in the next subsection.

\subsubsection{Calculating $\alpha$ and $\bar \kappa$}

The fluctuations we are interested in to obtain the transport coefficients $\alpha$ and $\bar \kappa$ in this geometry read
\bea
\delta ds^2&=&2(t \delta f_2(r)+\delta g_{tx})dt dx+2r^2H(r)^2\delta h_{rx}drdx\,,\nn \\
\delta A&=&(-t \delta f_1(r)+\delta a_x(r))dx\,.
\eea

The explicit form of the electric and heat current that do not depend on the radial coordinate reads
\bea
J&=&\sqrt{\frac{11}{2}}\frac{r^{3/2}}{N}\left(\delta g_{tx}A_t'+r^3F N^2\delta a_x'+t(\delta f_2 A_t'-r^3F N^2\delta f_1')\right)\,,\nn\\
Q&=&J A_t-\sqrt{\frac{11}{8}}F^2N^3r^{\frac{15}{2}}\left[t\left(\frac{\delta f_2}{N^2F r^3}\right)'+\left(\frac{\delta g_{tx}}{N^2F r^3}\right)'\right],
\eea
and using the following choice
\bea
\delta f_1&=&E+\zeta A_t \,, \nn\\
\delta f_2&=& \zeta r^3 N^2 F \,,
\eea
we can cancel the terms with temporal dependence.

The remaining Einstein equation reads
\be 
\delta h_{rx}=\frac{2 H \left(H \left(r \delta f_2'-2 \left(r \delta f_1 A_t'+\delta f_2\right)\right)-2 r \delta f_2
   H'\right)}{11 r^6 F N^2}
\ee
Because $J$ and $Q$ are constants we can evaluate it on the horizon. In order to do that we use the Eddington-Filkenstein coordinates \eqref{Edding} to obtain the near horizon behavior of $\delta g_{tx}$ and $\delta A_x$ ensuring that the fields are regular at $r_h$
\bea
\delta A_x&\sim & \frac{E}{r_h^{5/2}n_0^2 f_1}\log(r-r_h)\, ,\nn \\
\delta g_{tx}&\sim & -\left(r^2H^2\delta g_{rx}\right)_{_{r\rightarrow r_h}}-\zeta\frac{r^3 N^2F}{r_h^{5/2}n_0 f_1}\log(r-r_h)\,.
\eea
Using this is easy to obtain the values of the conserved $J$ and $Q$:
\bea
J&=&\frac{E  \left(-132 n_0^3 r_h^{7/2}+48 a_{t_1}^2 h_0^4\right)+a_{t_1} \zeta  r_h \left(-33 \left(3 h_0^4+1\right)
   n_0^2 r_h+4 a_{t_1}^2 h_0^4\right)}{12 \sqrt{22} n_0^3 r_h^{3/2}}\,, \nn \\
   Q&=&\frac{\left(4 a_{t_1}^2 h_0^4-33 \left(3 h_0^4+1\right) n_0^2 r_h\right) \left(\zeta \left(-33 \left(3 h_0^4+1\right)
   n_0^2 r_h+4 a_{t_1}^2 h_0^4\right) r_h +48 a_{t_1} h_0^4 E \right)}{576 \sqrt{22r_h} h_0^4 n_0^3 }.
\eea
The transport coefficients are defined by
\bea
\alpha &=&\frac{1}{T}\frac{\partial}{\partial \zeta}J=\frac{4 \sqrt{\frac{2}{11}} a_{t_1} h_0^4 \left(4 a_{t_1}^2 h_0^4-33 \left(3 h_0^4+1\right) n_0^2 r_h\right)}{\pi  n_0^2
  r_h^{3/2} \left(33 \left(3 h_0^4+1\right) n_0^2 r_h-a_{t_1}^2 h_0^4\right)}\,, \nn\\
\bar\kappa &=&\frac{1}{T}\frac{\partial}{\partial \zeta}Q=\frac{\left(4 a_{t_1}^2 h_0^4-33 \left(3 h_0^4+1\right) n_0^2 r_h\right)^2}{6 \sqrt{22r_h} \pi  n_0^2 \left(33
   \left(3 h_0^4+1\right) n_0^2 r_h-a_{t_1}^2 h_0^4\right)}.
\eea

Again we check  that  $\alpha=\bar\alpha$.
The thermal conductivity at zero electric current
\be 
\kappa=\bar\kappa-\frac{\alpha \bar\alpha T}{\sigma}=-\frac{\sqrt{\frac{11}{2}} n_0 r_h^3 \left(4 a_{t_1}^2 h_0^4-33 \left(3 h_0^4+1\right) n_0^2 r_h\right)^2}{6 \pi 
   \left(a_{t_1}^2 h_0^4-33 \left(3 h_0^4+1\right) n_0^2 r_h\right) \left(4 a_{t_1}^2 h_0^4+11n_0^3
   r_h^{7/2}\right)}\,.
\ee
\begin{figure}[!h]
\begin{center}  
\includegraphics[scale=0.65]{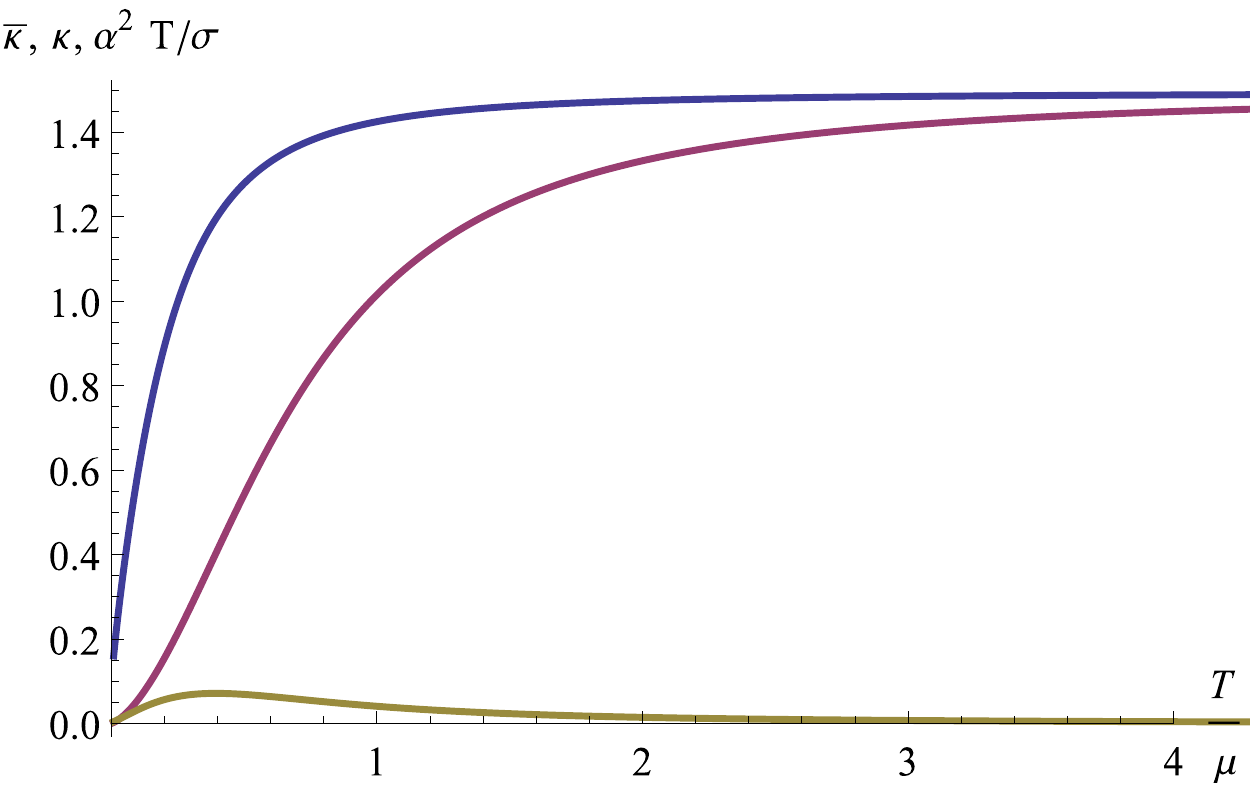}\hfill \includegraphics[scale=0.65]{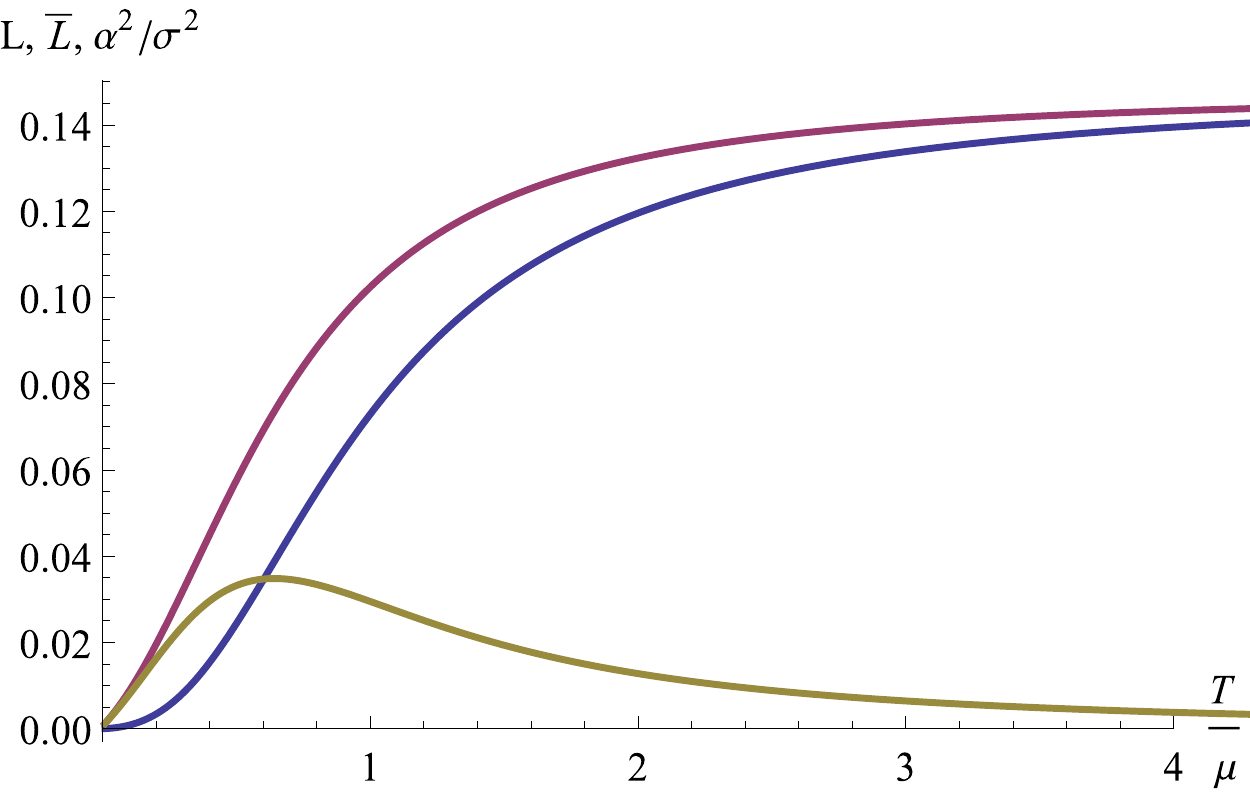}
\caption{Left: Thermal conductivities $ \bar \kappa$ (Blue) and $ \kappa$ (Purple) and the difference between the two of them $\alpha^2 T/\sigma$ (Yellow) as a function of the reduced temperature. Right:  Wiedemann-Franz ratios $L$ (Blue) and $\bar L$ (Purple) and the difference between these two $\alpha^2/\sigma^2$ (Yellow) as a function of the temperature. \label{fig3b}}
\end{center}  
\end{figure}

In the left hand side panel of Figure \ref{fig3b} we present the numerical results for the thermal conductivities  $ \bar \kappa$ and $ \kappa$. We find that these coefficients behave as $\bar \kappa\sim T^{1/2}$ and $\kappa\sim T$ at low temperatures. As we increase the temperature we find a crossover to a high temperature regime with $\kappa\sim \bar \kappa \gg  \alpha^2 T/\sigma $. 

From this analisis we could say that our charged Nil black hole interpolates between a low $T$ regime characterized by $\bar \kappa \gg \kappa$ to a large $T$ one with  $\kappa\sim \bar \kappa$. For a fermionic system, a $\bar \kappa \gg \kappa$ regime was associated in \cite{Mahajan:2013cja} to a hydrodynamic non-Fermi liquid, while $\kappa\sim \bar \kappa  \gg  \alpha^2 T/\sigma$ is usually associated to long-lived quasi-particles.

Also of particular interest are the  Wiedemann-Franz ratios
\bea
\bar L&=&\frac{\bar\kappa}{\sigma T}=\frac{16 h_0^4 \left(4 a_{t_1}^2 h_0^4 n_0-33 \left(3 h_0^4+1\right) n_0^3 r_h\right)^2}{\left(\pi  a_{t_1}^2
  h_0^4-33 \pi  \left(3 h_0^4+1\right)n_0^2 r_h\right)^2 \left(4 a_{t_1}^2 h_0^4+11 n_0^3 r_h^{7/2}\right)}\,, \nn\\
L &=&\frac{\kappa}{\sigma T}=\frac{176 h_0^4 n_0^5 r_h^{7/2} \left(4 a_{t_1}^2 h_0^4-33 \left(3 h_0^4+1\right) n_0^2 r_h\right)^2}{\pi ^2
   \left(a_{t_1}^2 h_0^4-33 \left(3 h_0^4+1\right) n_0^2 r_h\right)^2 \left(4 a_{t_1}^2 h_0^4+11 n_0^3
   r_h^{7/2}\right)^2}.
\eea
We plot these ratios in the right hand side panel of Figure \ref{fig3b} and can see that the law is broken at low temperatures and recovered as we increase $T$. The temperature scale at which the  Wiedemann-Franz law is recovered seems from the numerical data to coincide with the temperature at which the thermal conductivities $\kappa$ and $\bar \kappa$ saturate and also at which the resistivity $1/\sigma$ saturates, respecting the Mott-Ioffe-Regel bound.

\section{Charged nilgeometry black holes with hyperscaling violation}
\label{sec3}

In this section we will solve the Einstein - Maxwell equations without cosmological constant and show that there exists a charged Black hole with Nilgeometry in the near horizon and with hyperscaling violation \cite{Dong:2012se}.  Also, we compute the transport coefficients of the dual field theory and show that they can be obtained through the behavior of the fields on the horizon.


 %
%
%
%
%

\subsection{Solutions}
\label{sec31}

We will look now to a charged generalization to the Nil geometry black hole with hyperscaling violation presented in  \cite{Hassaine:2015ifa}. Hence we will consider the ansatz
\bea
A&=&A_t(r)\  dt\, ,\nn  \\ 
r^{\frac{2\theta}{3}}ds^2&=&- r^{3} N^2(r)F(r) dt^2 + \frac1{r^2 F(r)} dr^2 +r^2 H^2(r)\left( dx^2 +  dy^2\right)+ r^4\left(   dz-x dy   \right)^ 2\,,\label{hypernil}
\eea
with hyperscaling violation exponent $\theta=9/2$.

The equations of motion read
\bea
F'&=&\frac{-4 r H' \left(4 r^2 H^4 A_t'^2+3 N^2 \left(F H^4-2\right)\right)+10 r^2 H^5 A_t'^2-36 r^2 F H^3 N^2 H'^2+3 H
  N^2 \left(F H^4-1\right)}{3 r H^4 N^2 \left(4 r H'-H\right)},\nn\\
H''&=&\frac{H^3 \left(r H'-H\right) \left(2 r H A_t'^2+3 F N^2 H'\right)-3 N^2 H'}{3 r F H^4 N^2},\nn\\
N'&=&\frac{-2 H' \left(4 r^2 H^4 A_t'^2+3 N^2 \left(F H^4-2\right)\right)+8 r H^5 A_t'^2-12 r F H^3 N^2 H'^2}{3 F H^4
   N \left(H-4 r H'\right)},\nn\\
  A_t''&=&-\frac{2 A_t' \left(-4 r^3 H^4 A_t'^2 H'+4 r^2 H^5 A_t'^2-3 F H^3 N^2 \left(-2 r^2 H'^2-2 r H H'+H^2\right)+6 r
  N^2 H'\right)}{3 r F H^4 N^2 \left(4 r H'-H\right)}.
\eea

Near the horizon
\bea
A_t&=&a_{t_1}(r-r_h)-\frac{a_{t_1}}{r_h}(r-r_h)^2+\ldots,\nn\\
   F&=&\left(\frac{1}{h_0^4 r_h}-\frac{2 a_{t_1}^2 r_h}{3 n_0^2}\right)(r-r_h) +\ldots,\nn\\
H&=&h_0+\frac{2 \,a_{t_1}^2 h_0^5\, r_h}{2 a_{t_1}^2 h_0^4 r_h^2-3 n_0^2}(r-r_h)+\ldots\nn,\\
  N&=&n_0+\frac{4 a_{t_1}^2 h_0^4 n_0 r_h}{2 a_{t_1}^2 h_0^4 r_h^2-3 n_0^2}(r-r_h)\ldots.
\eea

We will shoot from the horizon towards the boundary looking for solutions that behave as
\bea
H(r)&\approx&1+\ldots\,,  \nn \\ \nn
F(r)&\approx&1-\frac{M}{r}+\ldots\,,\\ \nn
N(r)&\approx&1+\ldots\,,\\
A_0(r)&\approx&\mu-\frac{\rho}{r}+\ldots\,.
\label{bdynilt}
\eea
This solutions correspond asymptotically to Nil geometries with anisotropic asymptotic scaling 
\bea
t \rightarrow \lambda^{3/2}\, t\,,\,\,\,\, r \rightarrow \lambda^{-1} r \,, \,\,\,\, x\rightarrow \lambda\, x\, , \,\,\,\, y\rightarrow \lambda\, y \,, \,\,\,\, z\rightarrow \lambda^2\, z \,, \,\,\,\,\,\,\, ds\rightarrow \lambda^{3/2} \,ds \,.
\eea
\begin{figure}[!h]
\begin{center}  
\includegraphics[scale=0.59]{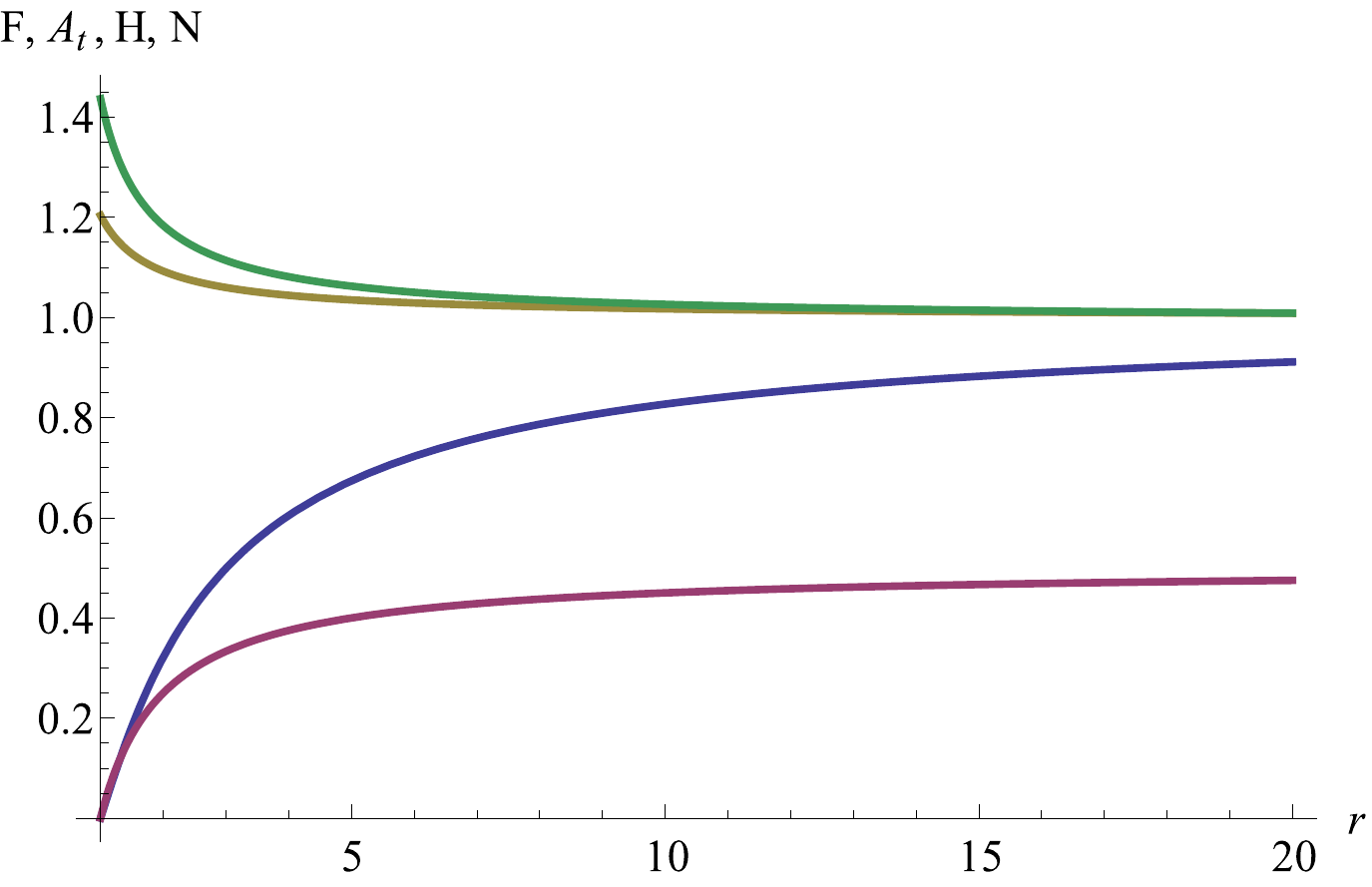}\hfill\includegraphics[scale=0.59]{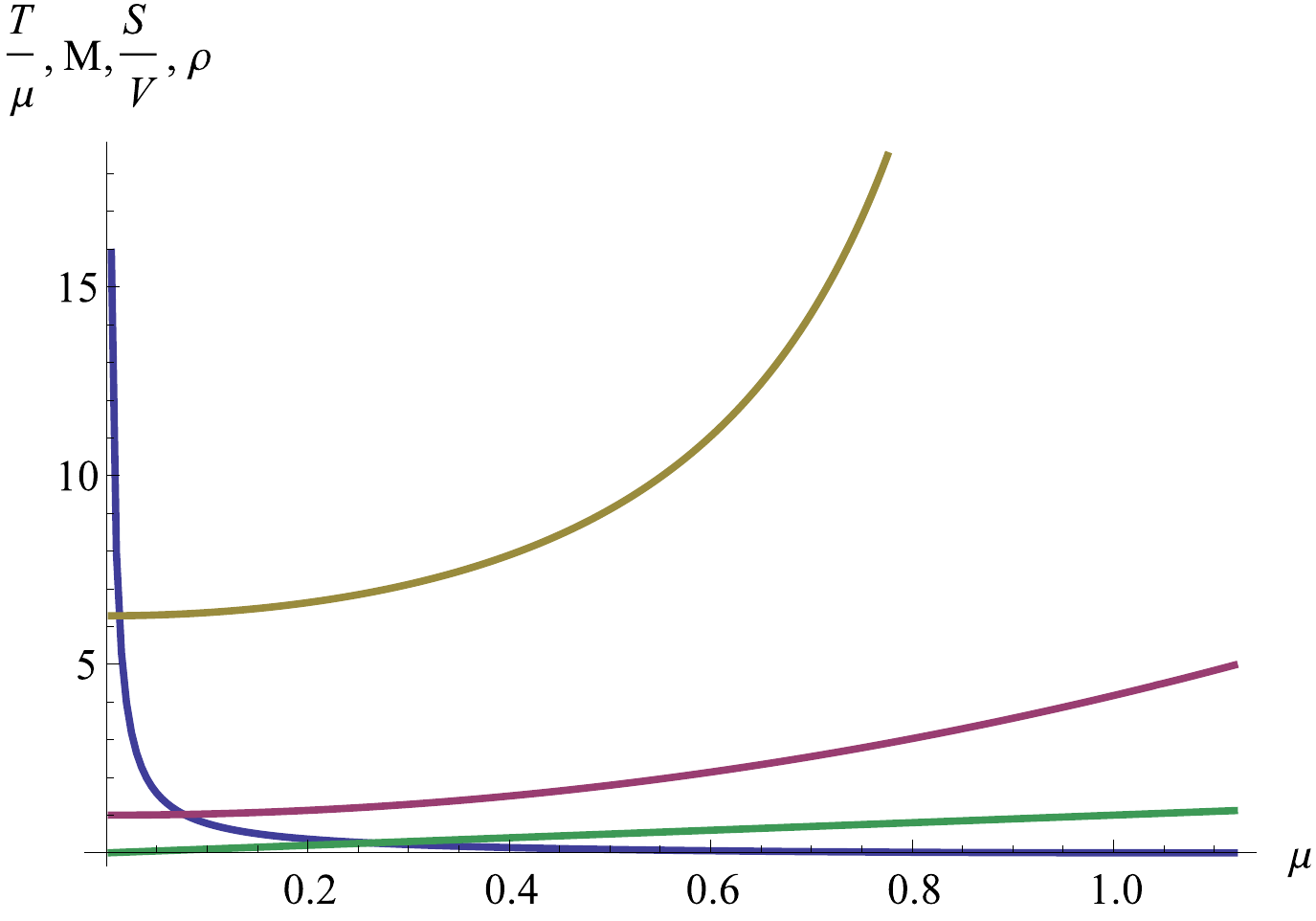}
\caption{Left: Radial profiles for the fields $F$ (Blue), $A_t$ (Purple), $H$ 
(Yellow) and $N$ (Green) for $\mu=1$. Right: Termperature $T$ (Blue), black hole mass $M$ (Purple), black hole entropy $S$ (Yellow) and  charge density $\rho$ (Green) for the dual field theory as a function of the chemical potential $\mu$.  \label{fig4}}
\end{center}  
\end{figure}

In the left panel of  Figure \ref{fig4} we show the profiles resulting from integrating the equations of motion. We use a shooting method with the horizon variables $h_0,\,n_0,\, a_{t_1}$  in order to get the desired boundary behavior (\ref{bdynilt}).
The temperature and entropy read
\bea
T&=&\frac{ n_0}{4\pi }\left(\frac1{r_hh_0^4}- \frac{2 r_h a_{t_1}^2}{3 n_0^2}     \right) \nn\,, \\ 
S&=& \frac{2 \pi h_0^2}{\sqrt{r_h}}   \,.
\eea
On the right hand side of  Figure \ref{fig4}, we show the dependence of the black hole's temperature, mass and entropy as well as the charge density as a function of the chemical potential of the dual field theory. 

\subsection{Finite conductivity from charged hyperscaling nilgeometry black holes}
\label{sec32}
We now proceed to compute the thermoelectric properties of Equation (\ref{ohm}) for this geometry.

\subsubsection{Calculating $\sigma$ and $\bar{\alpha}$}

In order to compute $\sigma$ and $\bar \alpha$ we will consider small perturbations around (\ref{hypernil}) of the form
\bea
\delta A&=&(-Et+\delta a_{x}(r)) dx\,, \nn \\ 
\delta ds^2&=& 2\  \delta g_{tx}(r)\  dt \, dx+ 2 r^2 H^2 \delta h_{rx}(r) \ dr \,dx\,.
\eea
The explicit expression for the electric current $J$ reads
\bea
J=\frac{\delta g_{tx} A_t'}{N}+r^3 F N \delta a_x'\,,\, 
\eea
and one can check that the Maxwell equation for $a_x$ is equivalent to $\partial_r J=0$, so we can evaluate $J$ at any $r$, including the black hole horizon.
\be 
Q=J A_t+\frac{3}{2}F^2r^3N^3\left(\frac{\delta g_{tx}}{r^3 N^2 F}\right)'\,.
\ee 
Again, asking for regularity of the fields on the horizon and using the coordinate $v$ defined on \eqref{Edding} we can evaluate explicitly the currents on the horizon

\bea 
J&=&E\left(\frac{2 a_{t_1}^2 h_0^4}{n_0^3 r_h^3}+\frac{8 r_h^3 \left(3 n_0^2-2 a_{t_1}^2 h_0^4
   r_h^2\right)}{33 \left(3 h_0^4+1\right) n_0^2 r_h-4 a_{t_1}^2 h_0^4}\right) \,,\nn\\
   Q&=&\frac{E \left(2 a_{t_1}^3 h_0^4 r_h^2-3 a_{t_1} n_0^2\right)}{3 n_0^3 r_h^4}\,,
\eea
and we can read the transport coefficients
\bea
\sigma &=&\frac{2 a_{t_1}^2 h_0^4}{n_0^3 r_h^3}+\frac{8 r_h^3 \left(3 n_0^2-2 a_{t_1}^2 h_0^4 r_h^2\right)}{33 \left(3
   h_0^4+1\right) n_0^2 r_h-4 a_{t_1}^2 h_0^4}\,, \nn\\
\bar\alpha &=&-\frac{4 \pi  a_{t_1} h_0^4}{n_0^2 r_h^3}\,,
\eea
\begin{figure}[!h]
\begin{center}  
\includegraphics[scale=0.6]{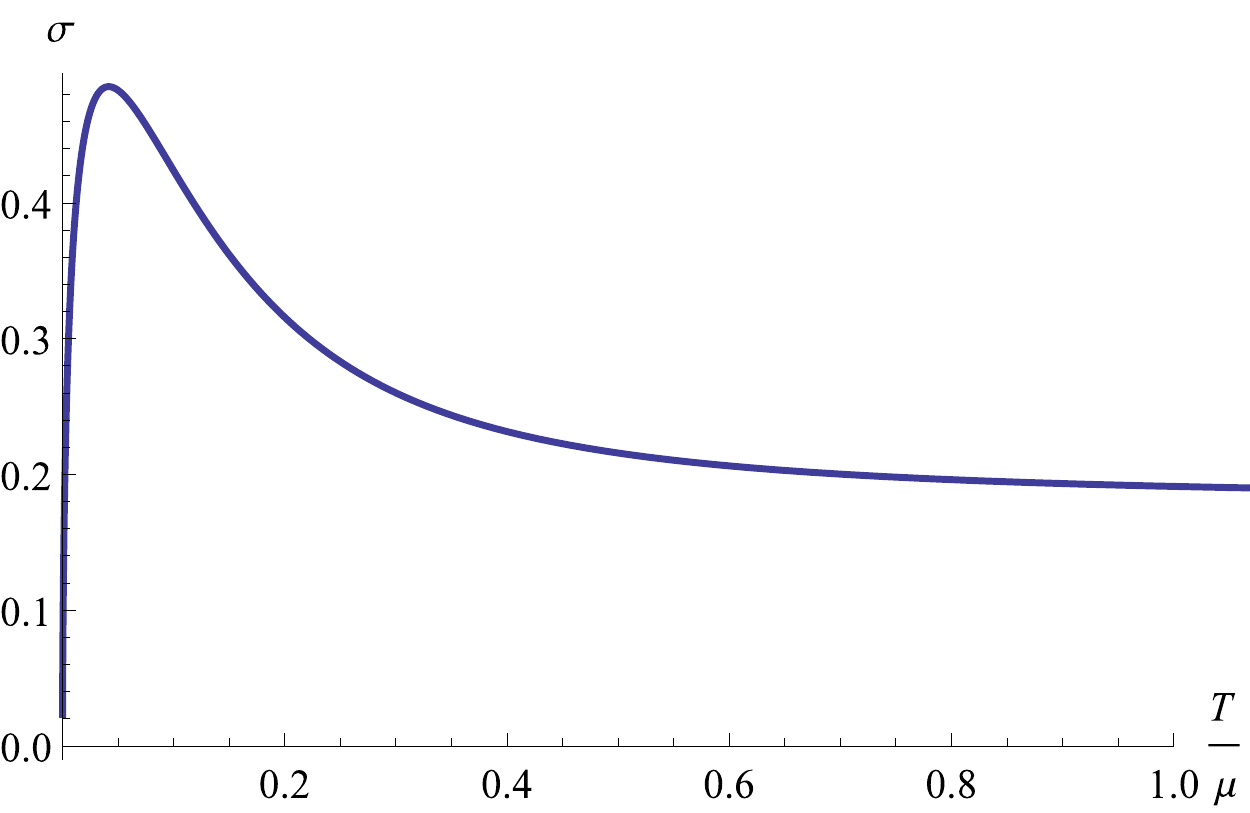}\hfill \includegraphics[scale=0.6]{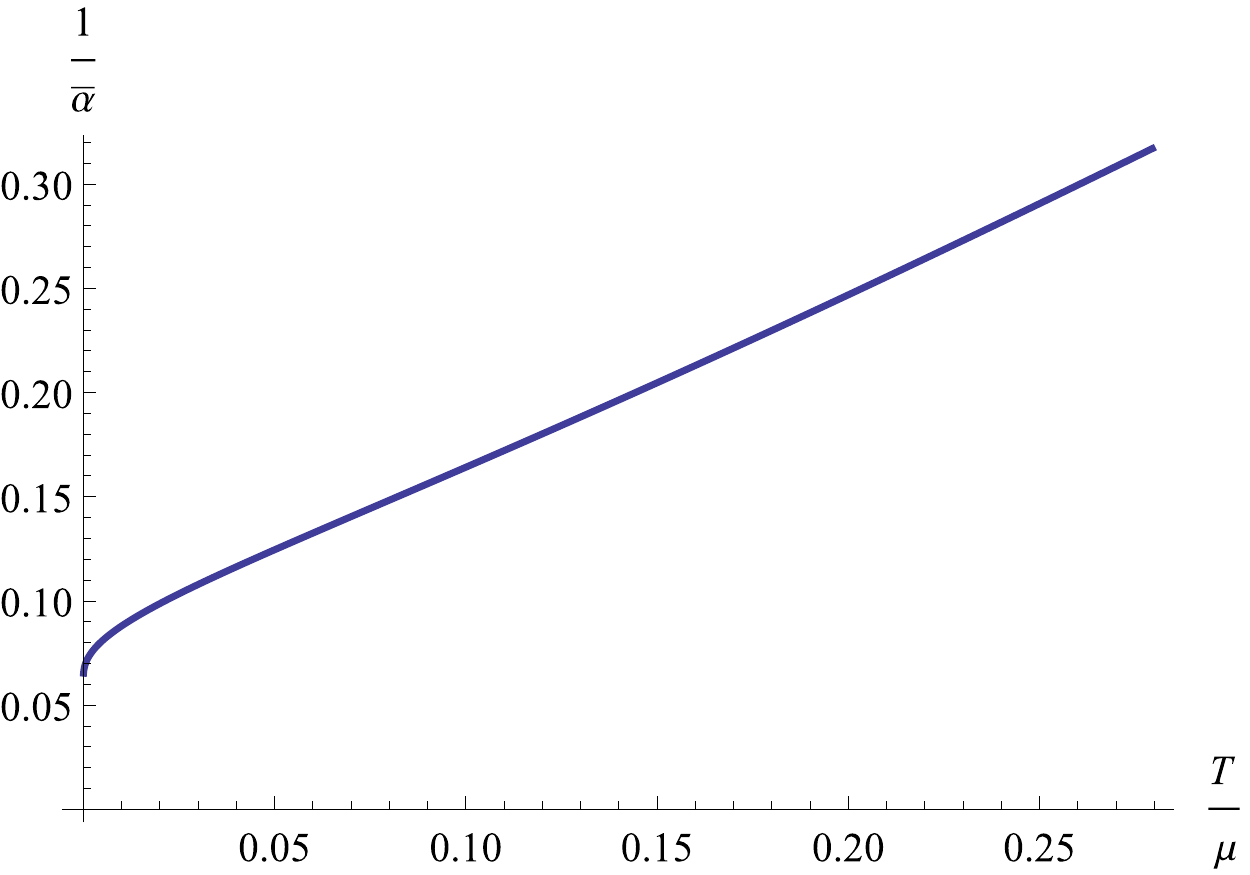}
\caption{Electric conductivity $\sigma$ and thermal resistivity $1/\bar\alpha$ as a function of the temperature.\label{fig5}}
\end{center}  
\end{figure}

In Figure \ref{fig5} we show the numerical computation of the DC conductivity $\sigma$  and the thermal resistivity and $1/\alpha$. We find that the resistivity diverges $1/\sigma \sim T^{-1/2}$ at low temperatures. This is a typical feature of an insulator \cite{Donos:2014uba}. As we increase the temperature  $1/\sigma$ decreases down to a minimum and then increases again. Finally it saturates at high $T$, indicating the presence of some sort of strongly interacting Mott-Ioffe-Regel bound. $1/\alpha$ on the other hand, seems finite in the $T=0$ limit, and behaves linear in $T$ for high enough temperatures.

\subsubsection{Calculating $\alpha$ and $\bar \kappa$}

We will now turn into the study of the fluctuations
\bea
\delta ds^2&=&2\left(t \delta f_2(r)+\delta g_{tx}\right) dt dx+2r^2H(r)^2\delta h_{rx} dr dx\,,  \nn \\
\delta A&=&(-t \delta f_1(r)+\delta a_x(r)) dx\,.
\eea
Maxwell equation and one of the Einstein's equations are equivalent to say that the following quantities are $r$ independent
\bea
J&=&\frac{1}{N}\left(\delta g_{tx} A_t'+r^3 F N^2 \delta a_x'+t(\delta f_2 A_t'-r^3 F N^2\delta f_1')\right), \nn\\
Q&=&J A_t+\frac{3}{2}F^2r^3N^3\left(\left(\frac{\delta g_{tx}}{r^3 N^2 F}\right)'+t\left(\frac{\delta f_2}{r^3 N^2 F}\right)'\right)\,.
\eea
The remaining Einstein equation reads
\bea
\delta g_{rx}=\frac{H \left(H \left(r\, \delta f_2'-2 \left(r^4 \delta f_1 A_t'+\delta f_2\right)\right)-2 r \delta f_2
   H'\right)}{r^6 F N^2}\,.
\eea
Using the following choice
 \bea
\delta f_1&=&E+\zeta A_t   \,,\nn\\
\delta f_2&=& \zeta r^3 N^2 F\,,
\eea
we can eliminate the temporal dependence. Using Eddington - Filkenstein coordinates \eqref{Edding} we can obtain for the near horizon 
\bea
\delta a_x&\sim & \frac{E}{r_h n_0^2 f_1}\log(r-r_h)\,,\nn\\
\delta g_{tx}&\sim & -\left(r^2H^2\delta g_{rx}\right)_{_{r\rightarrow r_h}}-\zeta\frac{r^3 N^2F}{r_h n_0 f_1}\log(r-r_h),
\eea
 and with this expressions we are prepared to compute the constants $J$ and $Q$ at $r_h$,
 
\bea
J&=& \frac{E}{n_0}  \left(\frac{2 a_{t_1}^2 h_0^4}{n_0^2 r_h^3}+\frac{8 n_0 r_h^3 \left(2 a_{t_1}^2 h_0^4 r_h^2-3
  n_0^2\right)}{33 \left(3 h_0^4+1\right) n_0^2 r_h-4 a_{t_1}^2 h_0^4}\right)-\frac{\zeta}{n_0} 
   \left(\frac{a_{t_1}}{r_h^4}-\frac{2 a_{t_1}^3 h_0^4}{3 n_0^2 r_h^2}\right),\nn\\
   Q&=&\frac{\left(2 a_{t_1}^2 h_0^4 r_h^2-3 n_0^2\right) \left(6 a_{t_1} h_0^4 r_h E -\zeta  \left(3 n_0^2-2
   a_{t_1}^2 h_0^4 r_h^2\right)\right)}{18 h_0^4 n_0^3 r_h^5}\,.
\eea
Then, the transport coefficients read
\bea
\alpha &=&-\frac{4 \pi  a_{t_1} h_0^4}{n_0^2 r_h^3},\nn\\
\bar\kappa &=&\frac{2 \pi}{3 r_h^4}  \left(3-\frac{2 a_{t_1}^2 h_0^4 r_h^2}{n_0^2}\right).
\eea
Again we can define the thermal conductivity at zero electric current

\be 
\kappa=-\frac{8 \pi  n_0 \left(3 n_0^2 r_h-2 a_{t_1}^2 h_0^4 r_h^3\right)^2}{3 \left(4 a_{t_1}^4 h_0^8+a_{t_1}^2
   h_0^4 n_0^2 r_h \left(-99 h_0^4+8 n_0 r_h^7-33\right)-12 n_0^5 r_h^6\right)}.
\ee
\begin{figure}[!h]
\begin{center}  
\includegraphics[scale=0.64]{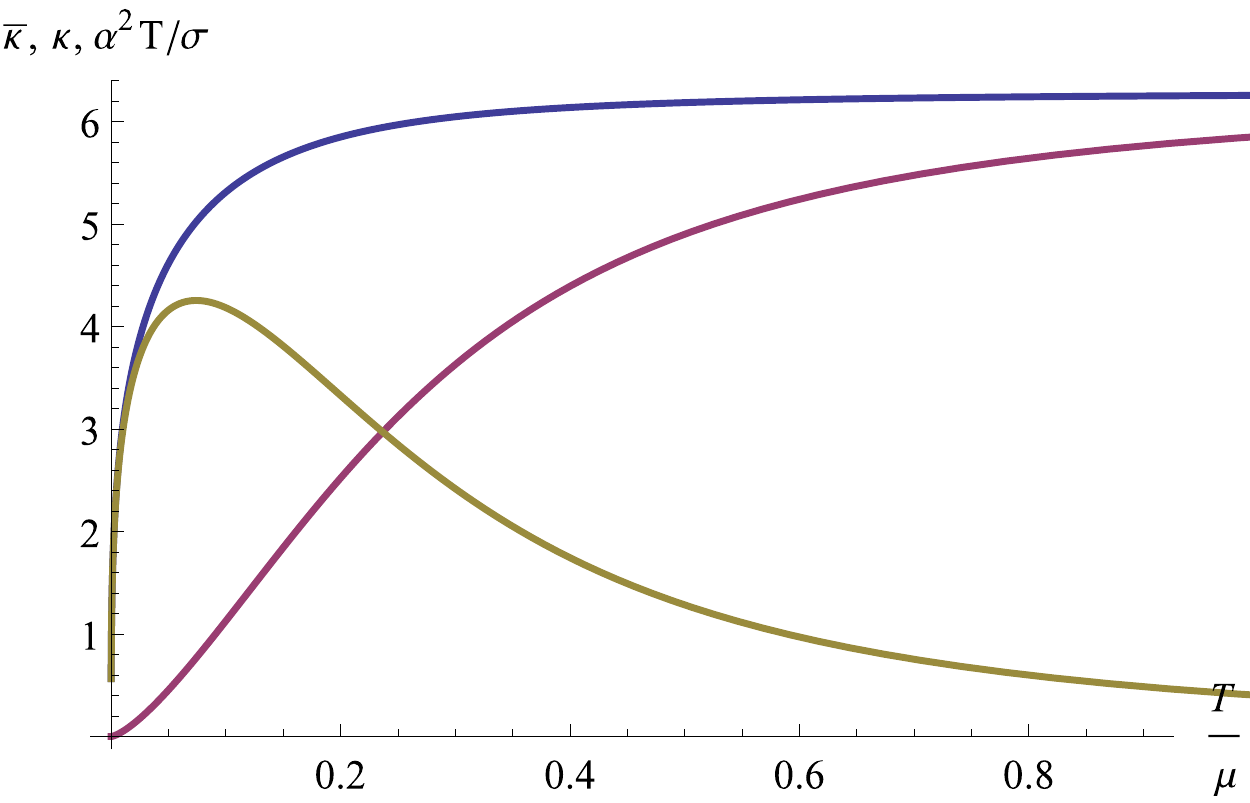}\hfill\includegraphics[scale=0.64]{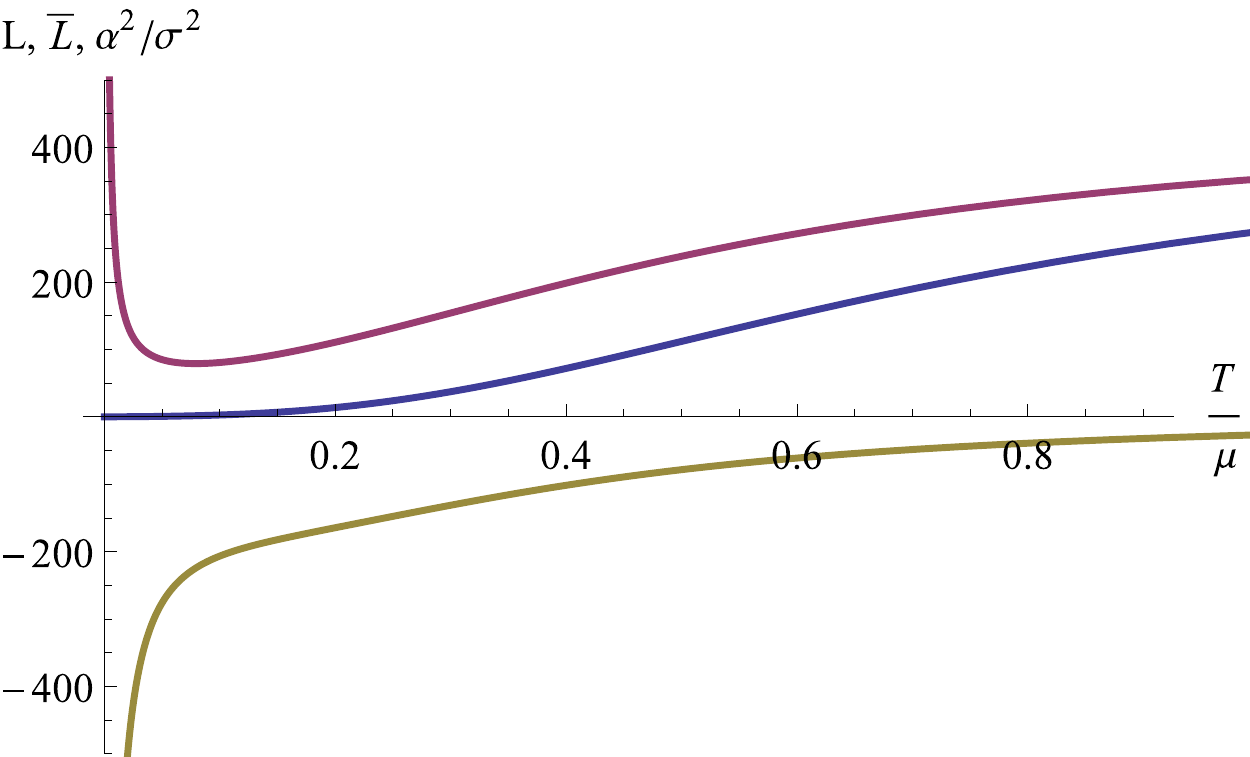}
\caption{Left: Thermal conductivities $ \bar \kappa$ (Blue) and $ \kappa$ (Purple) and the difference between the two of them $\alpha^2 T/\sigma$ (Yellow) as a function of the reduced temperature. Right:  Wiedemann-Franz ratios $L$  (Blue) and $\bar L$ (Purple)  and the difference between these two $\alpha^2/\sigma^2$ (Yellow)  as a function of the temperature.  \label{fig6}}
\end{center}  
\end{figure}

On the left hand side of Figure \ref{fig6} we plot the numerical values for the thermal conductivities $ \bar \kappa$ and $ \kappa$ as a function of the reduced temperature. We find that at low $T$, $ \bar \kappa \sim T^{1/3}$ and $ \kappa \sim T^{4/3}$. This implies that  in the low temperature regime we have $\bar \kappa \sim \alpha^2 T/\sigma \gg \kappa$. At large temperatures both  $ \bar \kappa$ and $ \kappa$ saturate to the same constant giving  $\bar \kappa \sim  \kappa  \gg   \alpha^2 T/\sigma$.

Of particular interest are the ratios
\bea
L&=&-\frac{16 \pi ^2 h_0^4 n_0^5 r_h^6 \left(33 \left(3 h_0^4+1\right) n_0^2 r_h-4 a_{t_1}^2 h_0^4\right) \left(2
  a_{t_1}^2 h_0^4 r_h^2-3 n_0^2\right)}{\left(4 a_{t_1}^4 h_0^8+a_{t_1}^2 h_0^4 n_0^2 r_h \left(-99
  h_0^4+8 n_0 r_h^7-33\right)-12 n_0^5 r_h^6\right)^2}\,, \nn \\
\bar L &=&\frac{4 \pi ^2 h_0^4 n_0^2 \left(33 \left(3 h_0^4+1\right) n_0^2 r_h-4 a_{t_1}^2 h_0^4\right)}{-4 a_{t_1}^4
   h_0^8+a_{t_1}^2 h_0^4 n_0^2 r_h \left(99 h_0^4-8 n_0 r_h^7+33\right)+12 n_0^5 r_h^6}\,,
\eea
where $L$ gives the Wiedemann-Franz law, and is a constant for systems where charge carriers are the responsible for heat transport. Since our system in an insulator and conductivity goes to zero at low temperatures, we observe that $ \bar L$ is divergent at low $T$. On the other hand, $ L$ is finite in the hole range of temperatures and saturares to a constant at high $T$.

\section{Conclusions}

Along this work we solve the Einstein - Maxwell equations of motion asking for charged black hole solutions with Solv, Nil and hyperscaling Nil geometries.

We compute the conductivity matrix for the dual field theory to these geometries using the horizon data. We compared our results with the expectations for fermionic systems at finite chemical potential following recent results presented in \cite{Mahajan:2013cja}. This comparison might be a bit loose, in the sense that we do not know the matter content of the dual field theory. Since having a Fermi surface is a natural way to have massless exitations, one might expect that the fermionic dergees of freedom of the dual field theory govern the low energy dynamics. Hence, we can hope that this comparison makes sense.

We found that the Solv and Nil geometries are dual to strongly coupled  metals. 
Remarkably we observe that the  Mott-Ioffe-Regel bound is satisfied even in these strongly coupled systems in the sense that resistivities go to a constant at high temperatures.  Moreover, at high temperatures $\kappa \sim\bar\kappa  \gg  \alpha^2 T/\sigma$ which suggest that the heat and electric current are carried by quasiparticles in this regime. At low temperatures we find different qualitative behaviors for these geometries, since the resistivity goes to zero for the Nil geometry while it remains finite for the Solv geometry.
On the other hand, from the transport coefficients computed from the near horizon behaviour of the charged hyperscaling Nilgeometry we conclude that the dual field theory describes an insulator.

Let us now conclude by discussing some possible open directions.
One interesting open direction would be to thoroughly study the behavior of these metrics with magnetic field expanding on our results of Section \ref{sec11}. Interesting phenomenology may be obtained from the study of transport coefficients in the presence of magnetic fields and some work has been done trying to reproduce the cuprates phenomenology from momentum dissipating holography \cite{Blake:2014yla,Blake:2015ina, Amoretti:2016cad} . Furthermore, for our self dual solutions in the sense defined in Equation (\ref{duality}) one might expect interesesting constraints for the transport coefficients \cite{Donos:2017mhp} coming from the enhanced symmetry and the fact that in some sense the system behave as living in a mixed dimension \cite{Hsiao:2017lch}.

Another interesting direction would be to study hairy black hole solutions which are dual to superfluid solutions. For this we would need to add some extra matter content. Classical examples are a charged scalar field \cite{Hartnoll:2008kx} or a charged Proca field  \cite{Cai:2013aca} or a $SU(2)$ Yang-Mills action \cite{Winstanley:2008ac,Gubser:2008zu}. In this direction it was found that the condensation of a scalar field restores the isotropy of the Bianchi $VII_0$ helical lattice \cite{Erdmenger:2015qqa}. It would be interesting to analize if this is indeed the case for our solutions.

\subsection*{Acknowledgments}
We would like to thank Julio Oliva for triggering this project. Particular thanks go to Daniele Musso for feedback on this manuscript.  We also thank ICTP-SAIFR school on AdS/CMT. ISL would like to thank MVC for hospitality.
 The authors are supported by CONICET.

\end{document}